%% file: aipsamp.tex
\documentclass[%
 aip,
 amsmath,amssymb,
 reprint,%
]{revtex4-1}

\usepackage{graphicx}
\usepackage{dcolumn}
\usepackage{bm}

\usepackage[utf8]{inputenc}
\usepackage[T1]{fontenc}
\usepackage{mathptmx}
\usepackage{braket}
\usepackage{color}
\usepackage{xcolor}
\usepackage{qcircuit}
\usepackage[caption=false]{subfig}
\usepackage{hyperref}
\def\ketbra#1#2{{\vert#1\rangle\!\langle#2\vert}} 
\usepackage{subfiles}

\begin{document}

\preprint{AIP/123-QED}

\title{Variational Simulation of Schwinger's Hamiltonian\\ with Polarisation Qubits}

\author{O.\,V.~Borzenkova}\email{oksana.borzenkova@skoltech.ru}
\affiliation{Skolkovo Institute of Science and Technology, 3 Nobel Street, Moscow 121205, Russian Federation}
\author{G.\,I.~Struchalin}
\affiliation{Quantum Technology Centre and Faculty of Physics,  M.V.~Lomonosov Moscow State University,  1 Leninskie Gory Street, Moscow 119991, Russian Federation} 
\author{A.\,S.~Kardashin}
\affiliation{Skolkovo Institute of Science and Technology, 3 Nobel Street, Moscow 121205, Russian Federation}
\author{V.\,V.~Krasnikov}
\affiliation{Quantum Technology Centre and Faculty of Physics,  M.V.~Lomonosov Moscow State University,  1 Leninskie Gory Street, Moscow 119991, Russian Federation} 
\author{N.\,N.~Skryabin}
\affiliation{Quantum Technology Centre and Faculty of Physics,  M.V.~Lomonosov Moscow State University,  1 Leninskie Gory Street, Moscow 119991, Russian Federation} 
\author{S.\,S.~Straupe}
\affiliation{Quantum Technology Centre and Faculty of Physics,  M.V.~Lomonosov Moscow State University,  1 Leninskie Gory Street, Moscow 119991, Russian Federation} 
\author{S.\,P.~Kulik}\homepage{https://quantum.msu.ru}
\affiliation{Quantum Technology Centre and Faculty of Physics,  M.V.~Lomonosov Moscow State University,  1 Leninskie Gory Street, Moscow 119991, Russian Federation} 
\author{J.\,D.~Biamonte}\homepage{https://quantum.skoltech.ru}
\affiliation{Skolkovo Institute of Science and Technology, 3 Nobel Street, Moscow 121205, Russian Federation}

\date{\today}

\begin{abstract}
The numerical emulation of quantum physics and quantum chemistry often involves an intractable number of degrees of freedom and admits no known approximation in general form.
In practice, representing quantum-mechanical states using available numerical methods becomes exponentially more challenging with increasing system size.
Recently quantum algorithms implemented as variational models, have been proposed to accelerate such simulations.  
Here we study the effect of noise on the quantum phase transition in the Schwinger model, within a variational framework. 
The experiments are built using a free space optical scheme to realize a pair of polarization qubits and enable any two-qubit state to be experimentally prepared up to machine tolerance.  
We specifically exploit the possibility to engineer noise and decoherence for polarization qubits to explore the limits of variational algorithms for NISQ architectures in identifying and quantifying quantum phase transitions with noisy qubits.
We find that despite the presence of noise one can detect the phase transition of the Schwinger Hamiltonian even for a two-qubit system using variational quantum algorithms.
\end{abstract}

\maketitle

\begin{figure*}[t] 
\includegraphics[width=0.9\textwidth]{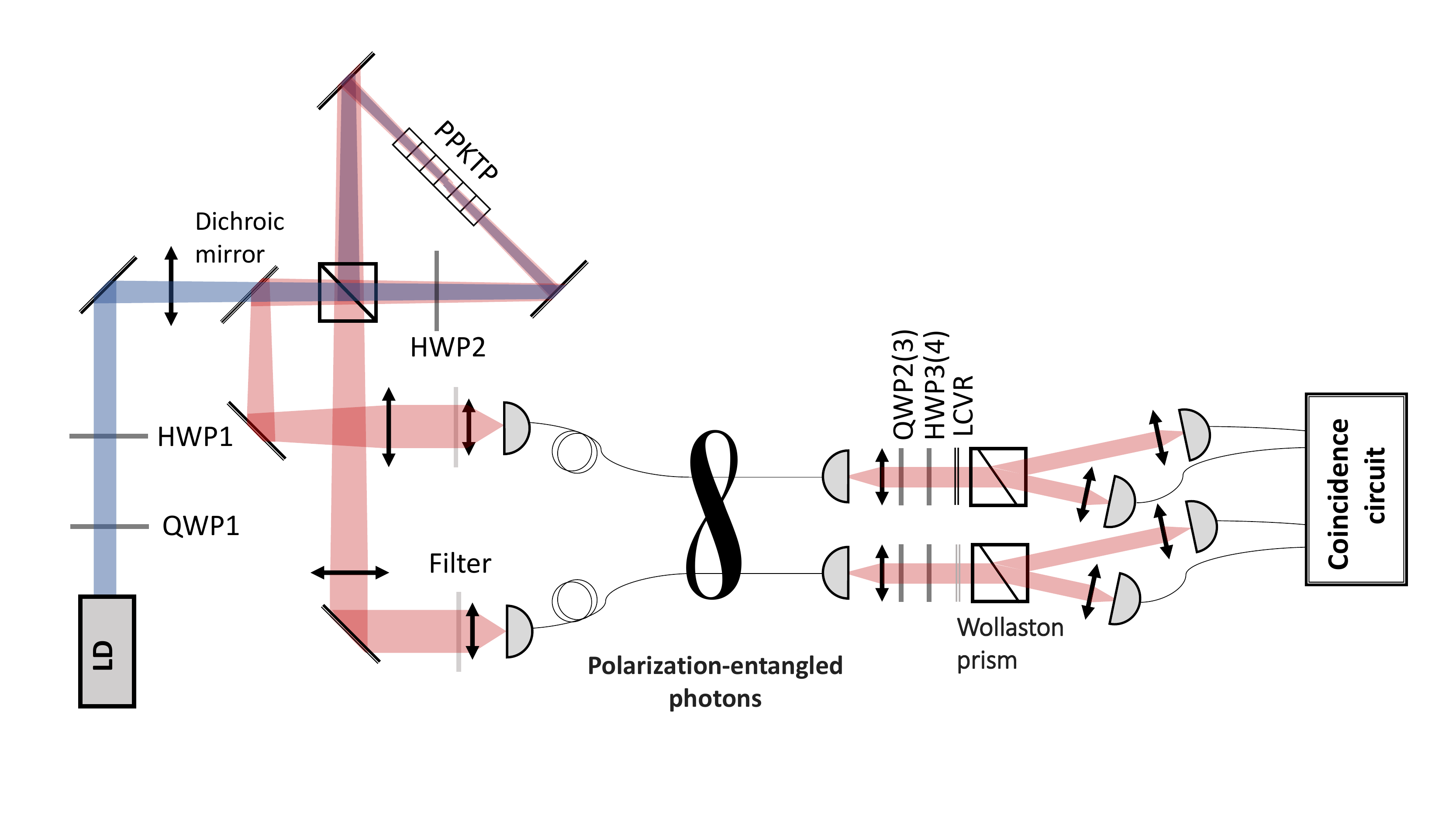}
\caption{\label{fig:epsart} Experimental setup implementing the variational quantum eigensolver algorithm using a pair of polarization qubits. Half-wave plates (HWP3, HWP4) and quarter-wave plates (QWP2, QWP3) in each channel prepare the desired variational state $|\psi(\boldsymbol\theta)\rangle$. Wollaston prisms implement projective measurements in the computational basis (horizontal and vertical polarization). HWP1 and QWP1 control the polarization of the pump beam, while HWP2 rotates the polarization by 90$^\circ$ to ensure the same polarization of the pump for the two pumping directions for the nonlinear crystal (PPKTP) inside the Sagnac interferometer. Liquid crystal variable retarders (LCVR) are used to artificially introduce dephasing noise on the first qubit, when necessary.}
\end{figure*}

The numerical emulation of quantum systems underpins a wide assortment of science and engineering and touches on fields ranging from statistical and quantum physics to biology and even to life- \cite{lambert2013quantum,neukart2017traffic} and behavioral-sciences \cite{werlang2010quantum, abadie2011gravitational,lubasch2020variational}. A physical simulator bootstraps one physical system to emulate the properties of another. While the time and memory required in the simulation of physical systems, particularly strongly correlated many-body quantum systems, using traditional computers often scales exponentially in the system size, the same is not always true for the physics-based quantum simulator.  Indeed, Richard Feynman first speculated that instead of viewing the simulation of quantum systems using classical computers as a no-go zone due to its apparent computational difficulty, Feynman argued~\cite{Feynman1986} that physical systems themselves naturally posses computational capacity to be harnessed and used.

Variational approaches to optimization and simulation of eigenstates \cite{obrien2014vqe,yung2014transistor, shen2017quantum, o2016scalable, kandala2017hardware, kokail2019self,Uvarov2020, wang2019accelerated} have been used recently to port ideas from machine learning \cite{2017Natur.549..195B} to enhance algorithms with quantum processors \cite{Akshay_2020,2017Natur.549..195B, farhi2014quantum,Uvarov2020, mitarai2019generalization}. These approaches rely on an iterative quantum-to-classical variational procedure. Proven to be a universal model of quantum computation in \cite{biamonte2019universal}---where the ansatz circuits are proven to be universal in \cite{morales2019universality}---the variational approach to quantum computation arose naturally as the pathway between a static simulator and a fully programmable gate-based quantum information processor. The variational model of quantum computation is the algorithmic workhorse of the current NISQ (Noisy Intermediate-Scale Quantum) technology era.

Recent experiments realize variational algorithms on different quantum hardware including superconducting qubits \cite{o2016scalable, kandala2017hardware}, trapped atoms \cite{yung2014transistor, shen2017quantum, kokail2019self} and photonic quantum processors \cite{obrien2014vqe,carolan2020variational}. The most common application of quantum variational algorithms includes quantum chemistry applications \cite{ryabinkin2018constrained, parrish2019quantum}. The original purpose of the algorithm was finding ground eigenvalues and eigenvectors; \cite{McClean_2018} shows that variational techniques can also find excited states, and various other proposals further expand the limits of applicability \cite{higgott2019variational}.

The variational quantum eigensolver (VQE) \cite{obrien2014vqe} performs classical optimization to minimize an expected Hamiltonian value. The purpose of this algorithm is to determine the eigenvalues of a particular Hamiltonian, which describes a physical system, for example, the interaction of spins or electronic systems \cite{yung2014transistor, shen2017quantum}. A classical computer initially sets a vector of parameters $\boldsymbol{\theta} = \{\theta_{i} \}$ for $i \in \mathbb{N}$ and an experimental setup prepares a parameterised quantum state $|\psi(\boldsymbol{\theta})\rangle$. After that, the state is measured, and the evaluation of the mean Hamiltonian value occurs. The parameters $\boldsymbol{\theta}$ are adjusted to find the ground-state energy:
\begin{equation}
E_{\mathrm{min}}(\boldsymbol{\theta}')= \min\limits_{\boldsymbol{\theta}} \langle\psi(\boldsymbol{\theta})|  H|\psi(\boldsymbol{\theta})\rangle.
\label{eq:Emin}
\end{equation}
Therefore, the problem consists of using classical optimization algorithms to select optimal parameters $\boldsymbol{\theta}$ corresponding to the (ideally) minimal value of energy (\ref{eq:Emin}).

Here we report an experimental implementation of VQE in a photonic system.  We target the exploration of a quantum phase transition in the Schwinger model. We specifically exploit the possibility to engineer noise and decoherence \cite{Pechen2011} for polarization qubits to explore the limits of variational algorithms for NISQ architectures in identifying and quantifying quantum phase transitions with noisy qubits. 

We implement VQE with polarization-encoded qubits using the experimental scheme shown in Fig.~\ref{fig:epsart}. Initial state preparation is carried out by a two-photon source based on spontaneous parametric down-conversion process (SPDC) in the Sagnac interferometer~\cite{zeilinger2007source}. A 405-nm laser diode beam is divided by a polarization beam-splitter (PBS), which makes it possible to pump a 30-mm long periodically poled potassium titanyl phosphate nonlinear (PPKTP) crystal in two opposite directions. As a result of a type-II SPDC, pairs of signal and idler photons with orthogonal polarizations are generated in both directions. Then each photon pair is divided on the PBS and sent to different arms of the scheme. Thus, at the output of the two-photon source, we have the following entangled state:
\begin{equation}
  |\psi_\text{in}\rangle=\alpha(\theta_1,\theta_2)|HV\rangle + \beta(\theta_1, \theta_2) |VH\rangle, \label{eq:SPDCOutputState}
\end{equation}
where the coefficients $\alpha$ and $\beta$ depend on the angular positions $\theta_1$ and $\theta_2$ of waveplates QWP1 and HWP1, which are placed in the pump beam. By rotating QWP1 and HWP1, we can alter the degree of entanglement of the initial state. The photon pairs are coupled to single-mode fibers and transferred to the measurement part of the setup. Motorized quarter-wave (QWP2, QWP3) and half-wave (HWP3, HWP4) plates are placed in each arm after the single-mode fiber channel, allowing to obtain any polarization state at the output. Finally, the Wollaston prism spatially separates the vertical and the horizontal polarizations to detect the prepared states using single-photon detectors in each of the arms. According to the measurement results, the classical algorithm transfers the new parameter values to the motorized plates until the optimal set of parameters is obtained.

We should note that estimation of a single mean value of a Hamiltonian requires projective measurements in several bases, while the Wollaston prism projects only onto $\ket{H}$ and $\ket{V}$ states. To change the basis one may use an additional pair of QWPs and HWPs, mounted just before the Wollaston prism. However, we chose a more economic setup, where the local unitary transformation of the initial state and the transformation of the measurement basis are compiled together, e.\,g.,
\begin{equation}
\bra{H} B U_\text{HWP}(\theta_4) U_\text{QWP}(\theta_3) = \bra{H} U_\text{HWP}(\theta'_4) U_\text{QWP}(\theta'_3), \label{eq:BasisChange}
\end{equation}
here $U(\theta_{3,4})$ is a transformation of a corresponding waveplate with an axis angles~$\theta_{3,4}$ and $B$ is a unitary matrix that changes the basis. New angles $\theta'_{3,4}$ are calculated automatically in our algorithm to perform measurements in desired bases.

\begin{figure}[tb]
    \centering
    \mbox{
        \Qcircuit @C=.5em @R=1.em {
            & \ket{0} & & \gate{U_\text{QWP}(\theta_1)} & \gate{U_\text{HWP}(\theta_2)} & \ctrl{1} & \gate{U_\text{QWP}(\theta_3)} & \gate{U_\text{HWP}(\theta_4)} & \meter \\
            & \ket{0} & & \gate{X}                      & \qw                           & \gate{X} & \gate{U_\text{QWP}(\theta_5)} & \gate{U_\text{HWP}(\theta_6)} & \meter
        }
    }
    \caption{\label{fig:gate_circuit} Schematic of the VQE algorithm. The initial state is prepared by three single-qubit gates and a Controlled-X gate. $U_\text{QWP}(\theta_1)$ and $U_\text{HWP}(\theta_2)$ are used to control the initial state in experiment. Other four single-qubit transformations serve both for the ansatz state preparation and the measurement basis change.}
\end{figure}
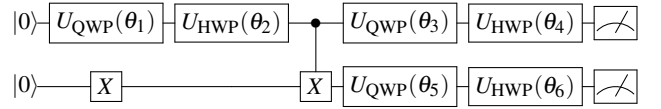

By mapping experimental optical elements to the gate model we arrive at the ansatz preparation circuit with six tunable parameters $\theta_i, i = 1, \dots, 6$ which is presented in Fig.~\ref{fig:gate_circuit}. The parameters $\theta_i$ physically correspond to the waveplates' rotation angles. A general waveplate with a phase shift $\delta$ and an axis position $\theta$ performs the transformation $U(\delta, \theta)$:
\begin{gather}
U(\delta, \theta) = V(\theta) D(\delta) V^\dagger(\theta),\label{eq:Waveplate}\\
V(\theta) = \openone \cos (\theta) - \imath \sigma^y \sin (\theta),  \quad
D(\delta) = e^{+\imath \delta \ket{1}\bra{1}}. \nonumber
\end{gather}
A controlled-X gate corresponds to SPDC in the nonlinear crystal.

Taking into account the ansatz preparation scheme, our VQE algorithm implementation consists of the four main steps:
\begin{enumerate}
    \item SPDC source emits the initial entangled state $\ket{\psi_\text{in}}$~\eqref{eq:SPDCOutputState}.
    \item Once the initial state has been prepared, a local unitary transformation $U_1 \otimes U_2$ is applied to get the probe state $\ket{\psi(\boldsymbol \theta)}$:
    \begin{equation}
        \ket{\psi(\boldsymbol \theta)} = U_1(\theta_3, \theta_4) \otimes U_2(\theta_5, \theta_6) \ket{\psi_\text{in}(\theta_1, \theta_2)}.
    \end{equation}
    Unitaries $U_1$ and $U_2$ are composed of the waveplate transformations: $U_1 = U_\text{HWP}(\theta_4) U_\text{QWP}(\theta_3)$, $U_2 = U_\text{HWP}(\theta_6) U_\text{QWP}(\theta_5)$.
    \item The cost function $E(\boldsymbol{\theta})=\langle\psi(\boldsymbol{\theta})|H|\psi(\boldsymbol{\theta})\rangle$ is calculated by summing up measurement results with coefficients depending on the problem Hamiltonian. Since usually Hamiltonian is expressed as a linear combination of Pauli observables and our setup allows only projective measurements, we first should decompose the Hamiltonian as a linear combination of projectors onto eigenbases of Pauli matrices. Change of basis is carried out according to the rule~\eqref{eq:BasisChange}.
    \item The value of $E(\boldsymbol{\theta})$ is minimized as a function of the parameters $\boldsymbol{\theta}$ using a classical optimizer routine. In particular we use simultaneous perturbation stochastic approximation (SPSA) algorithm.
\end{enumerate}


The Schwinger model describes interactions between Dirac fermions via photons in a two-dimensional space \cite{PhysRevD.66.013002, PhysRevA.73.022328}. In Ref.~\cite{kokail2019self}, the authors map the model to the lattice model of an electron-positron array. The Schwinger Hamiltonian exhibits a quantum phase transition: the signature of which (in finite dimensions) allows us to determine new features in VQE behavior and clarify its robustness to noise. 

The Schwinger Hamiltonian $H_N$ describes electron- positron pair creation and annihilation, their interaction and takes into account the particle mass:
\begin{equation}
 {H}_N = w \sum_{j=1}^{N-1}[ {\sigma}_j^+  {\sigma}_{j+1}^- + H.c.] + \frac{m}{2}\sum_{j=1}^N (-1)^j  {\sigma}_j^z + g \sum_{j=1}^N  {L}_j^2.
\end{equation}
It consists of the three terms: the first one is responsible for the interaction of an electron and a positron, the second depends on bare mass~$m$ of the particles, and the third stands for the energy of the electric field. We assume the coefficients $w = g = 1$ and only consider the dependence of the Hamiltonian ground energy on the bare mass. The operators in the third term are given by
\begin{equation}
 {L}_j = \epsilon_0 - \frac{1}{2} \sum_{l=1}^j [ {\sigma}_l^z +(-1)^l],
\end{equation}
where we set the background electric field parameter $\epsilon_0$ to zero.

The problem Hamiltonian can be encoded in the multiqubit system by using its decomposition into Pauli strings: $  P_{\alpha} =  \sigma_{1}^{\alpha_{1}} \otimes  \sigma_{2}^{\alpha_{2}} \otimes \ldots \otimes  \sigma_{N}^{\alpha_{N}}$ with single-qubit Pauli operators $ \sigma_{i}^{\alpha_{i}}\in \left\{I,  \sigma_{i}^{x},  \sigma_{i}^{y},  \sigma_{i}^{z}\right\}$ as
\begin{equation}
      H =\sum_a h_{\alpha}   P_{\alpha},
\end{equation}
where $N$ denotes the number of qubits and $h_\alpha \in \mathbb{R}$ are real coefficients. In further consideration, we will use this representation. We carried out numerical simulations and experiments for the case of two qubits, for which the Schwinger Hamiltonian takes the form
\begin{equation}
  H_2 = \openone +  \sigma_1^x \sigma_2^x +  \sigma_1^y \sigma_2^y - \frac{1}{2} \sigma_1^z + \frac{1}{2} \sigma_1^z \sigma_2^z + \frac{m}{2}( \sigma_2^z -  \sigma_1^z). \label{eq:H2}
\end{equation}

The quantum phase transition manifests itself in the behavior of the \emph{order parameter}
\begin{equation}
    \langle  {O} \rangle = \frac{1}{2N(N-1)} \sum_{j > i} \langle (1+(-1)^i {\sigma}_i^z)(1+(-1)^j {\sigma}_j^z) \rangle.
\end{equation}
For polarization-encoded pair of qubits the order parameter is simply a projector onto $\ket{VH}$ state:
\begin{equation}
    \langle  {O} \rangle = \frac{1}{4}\langle(1 -   \sigma_1^z)(1 +   \sigma_2^z)\rangle =  \langle \ketbra{VH}{VH} \rangle.
\end{equation}

\begin{figure}[tb]
    \centering
    \subfloat[]
    {
        \includegraphics[width=0.8\linewidth]{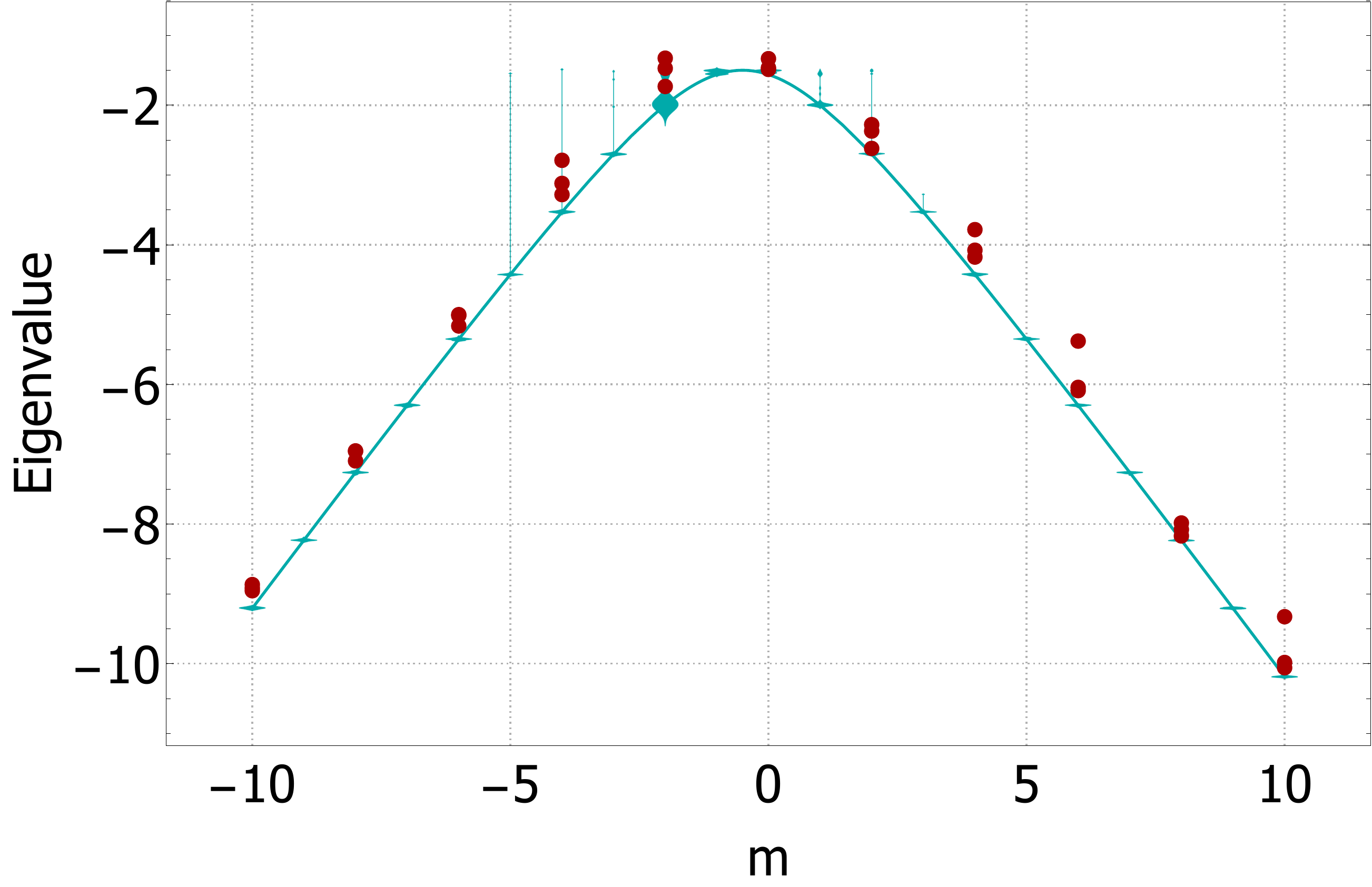}
        \label{fig:evall}
    }\\
    \subfloat[]
    {
        \includegraphics[width=0.8\linewidth]{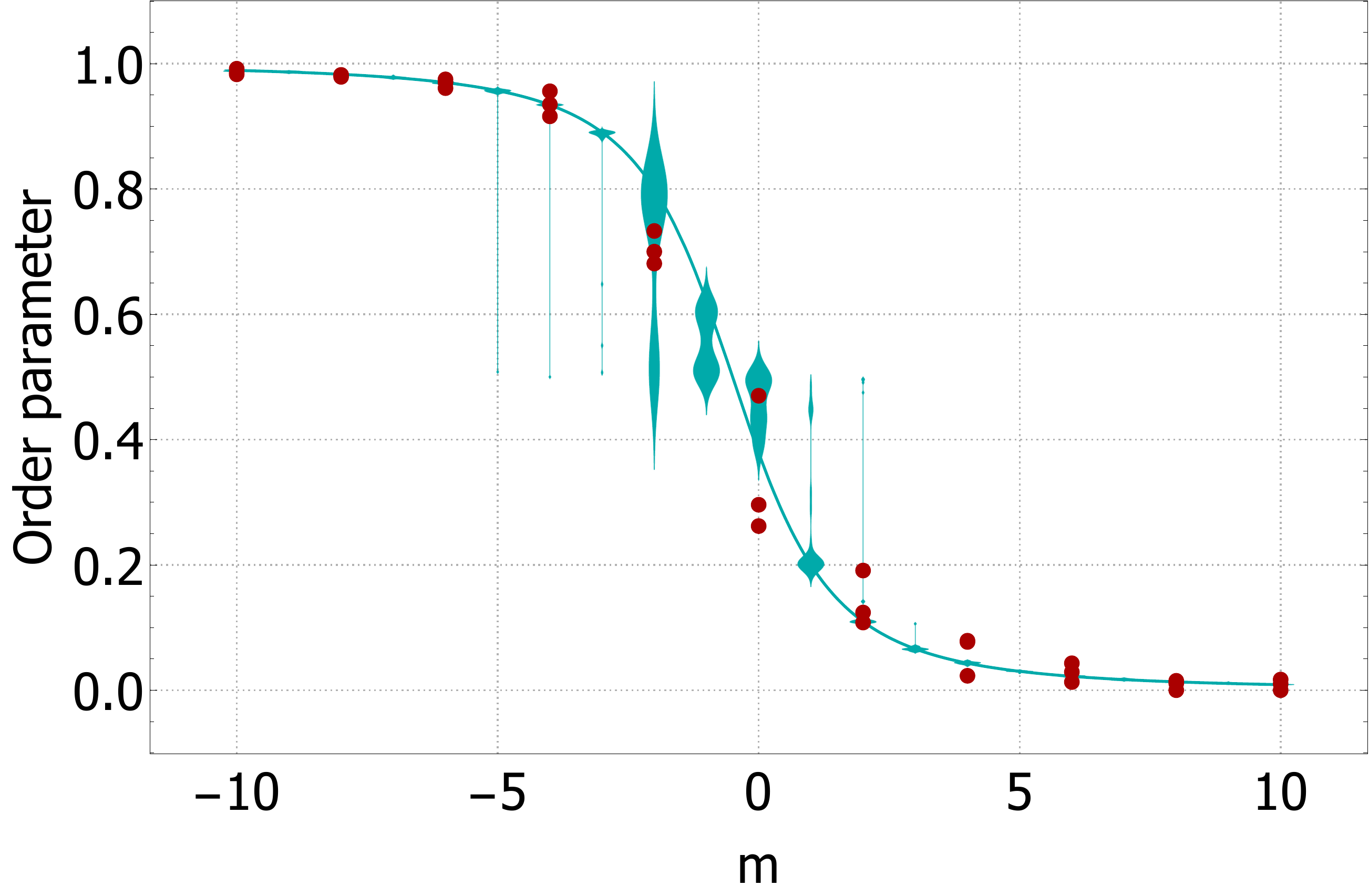}
        \label{fig:opall}
    }
    \caption{ Eigenvalue (a) and order parameter (b) versus bare mass $m$. Solid lines---analytical solution, cyan points (violin plot)---simulations, red points---experimental result. The probability distribution for each~$m$ in simulations is obtained using 30 trials. For the experiment all attempts are shown as distinct points.} 
    \label{fig:exp_no_noise}
\end{figure}

Two-qubit Schwinger Hamiltonian has four non-degenerate eigenvalues $E_1, \dots, E_4$. Two intermediate eigenvalues, $E_2 = 2$ and $E_3 = 1$, are constant and do not depend on the mass~$m$. The largest and the smallest eigenvalues, $E_{1,4} = 1/2 \pm \sqrt{m^2 + m + 17/4}$, vary with $m$ in a symmetric manner.

We are interested in the ground energy of the Hamiltonian that corresponds to the minimal eigenvalue $E_\text{min} \equiv E_4$. The graph of its dependence on $m$ is depicted in Fig.~\ref{fig:evall} and Fig.~\ref{fig:opall} shows the order parameter versus mass. The solid lines correspond to the exact analytical solutions, dots represent the results of simulations and experiment. A phase transition signifies itself in the rapid change of the order parameter from one to zero and it is expected near the point $m = -1/2$, where $\braket{  O} = 1/2$.

Exactly at the vicinity of the phase transition point $m = -1/2$, we found a discrepancy between analytical solutions and VQE simulations. The Hamiltonian $  H_2(m = -1/2)$ has the ground energy $E_\text{min} = -3/2$ with the corresponding eigenvector $(|01\rangle - |10\rangle)/\sqrt{2}$, which is a maximally entangled singlet state $\ket{\Psi^-}$.
A distinguishing feature of the singlet state is its invariance under local unitary rotations $U \in \text{SU}(2)$: $\ket{\Psi^-} = (U \otimes U) \ket{\Psi^-}$. Therefore, the target function $E(\boldsymbol \theta)$ remains constant on some parameter manifold. Note that this plateau does not change with the mass $m$, because $\forall m: \braket{\Psi^- |   H_2(m) | \Psi^-} = -3/2$.

For $m = -1/2$ the existence of the plateau is not a problem for the optimization algorithm, since the global minimum is attained at any point of the plateau. But for any $m \ne -1/2$ the minimum has a lower value, $E_\text{min} < -3/2$, while residing near the plateau with $E = -3/2$. So the landscape of $E(\boldsymbol \theta)$ in the punctured neighbourhood of $m = -1/2$ becomes a flat valley. A valley landscape puzzles the gradient-based optimizers and significantly slows convergence~\cite{Rosenbrock_CJ1960}. Therefore, the algorithm terminates at the wrong value. A little step noticeable in Fig.~\ref{fig:opall} illustrates this situation. When $m$ is far away from the phase transition point, the plateau does not strongly influence the results, because $E_\text{min}$ is much lower than $E = -3/2$.

In our particular case, slow convergence originated from the invariance of the singlet state~$\ket{\Psi^-}$ being the Hamiltonian eigenvector for $m = -1/2$. A more general view on the cause of the convergence problem is that it appears any time, when the ansatz is general enough to perform arbitrary local unitary transformations and the Hamiltonian ground state is close to some Bell state (not necessarily $\ket{\Psi^-}$). Indeed, all Bell states are equivalent under local transformations, so we can find a local map that brings a Bell state~$\ket{\psi_0}$ to a singlet one $\ket{\Psi^-}$:
\begin{equation}
    \ket{\Psi^-} = (W_1 \otimes W_2) \ket{\psi_0},
\end{equation}
where $W_{1,2}$ are some single-qubit unitary matrices. Consequently, an arbitrary Bell state $\ket{\psi_0}$ is invariant under the following transformation:
\begin{equation}
    \forall U \in \text{SU}(2): \ket{\psi_0} = (W_1^\dagger U W_1 \otimes W_2^\dagger U W_2) \ket{\psi_0}. \label{eq:BellInvariance}
\end{equation}
If ansatz circuit is general enough to prepare different transformations of the form~\eqref{eq:BellInvariance}, then the plateau in the landscape of $E(\boldsymbol\theta)$ appears. Therefore, when the Hamiltonian ground state is close to the Bell state, the nearby plateau will create flat valley landscape.

The simplest opportunity to get around poor optimizer convergence is by a correct choice of the initial point. We gathered statistics for $10^5$ random initial points~$\boldsymbol \theta$ for $m = -1/2$, $0$, $1/2$, and $10$ and found that near the phase transition the algorithm sticks to the plateau much frequently than to the proper minimum (see Supplementary material for details).

In order to clarify the issue with the accuracy, we used parameter $\delta$ from Ref.~\cite{bravo2020scaling}. This parameter characterises the closeness of the obtained energy $E$ to the exact ground level $E_0$ compared with the distance to the next energy level $E_1$: $\delta = \frac{E-E_0}{E_1-E_0}$. For ``good-enough'' accuracy, the parameter should be much less than one, $\delta \ll 1$. In our work, the maximum value of $\delta$ is $0.176$ for the experiment and $0.02$ for simulations.


\begin{figure*}[htb]
   \centering
   \subfloat[]
   {
        \includegraphics[width=0.42\textwidth]{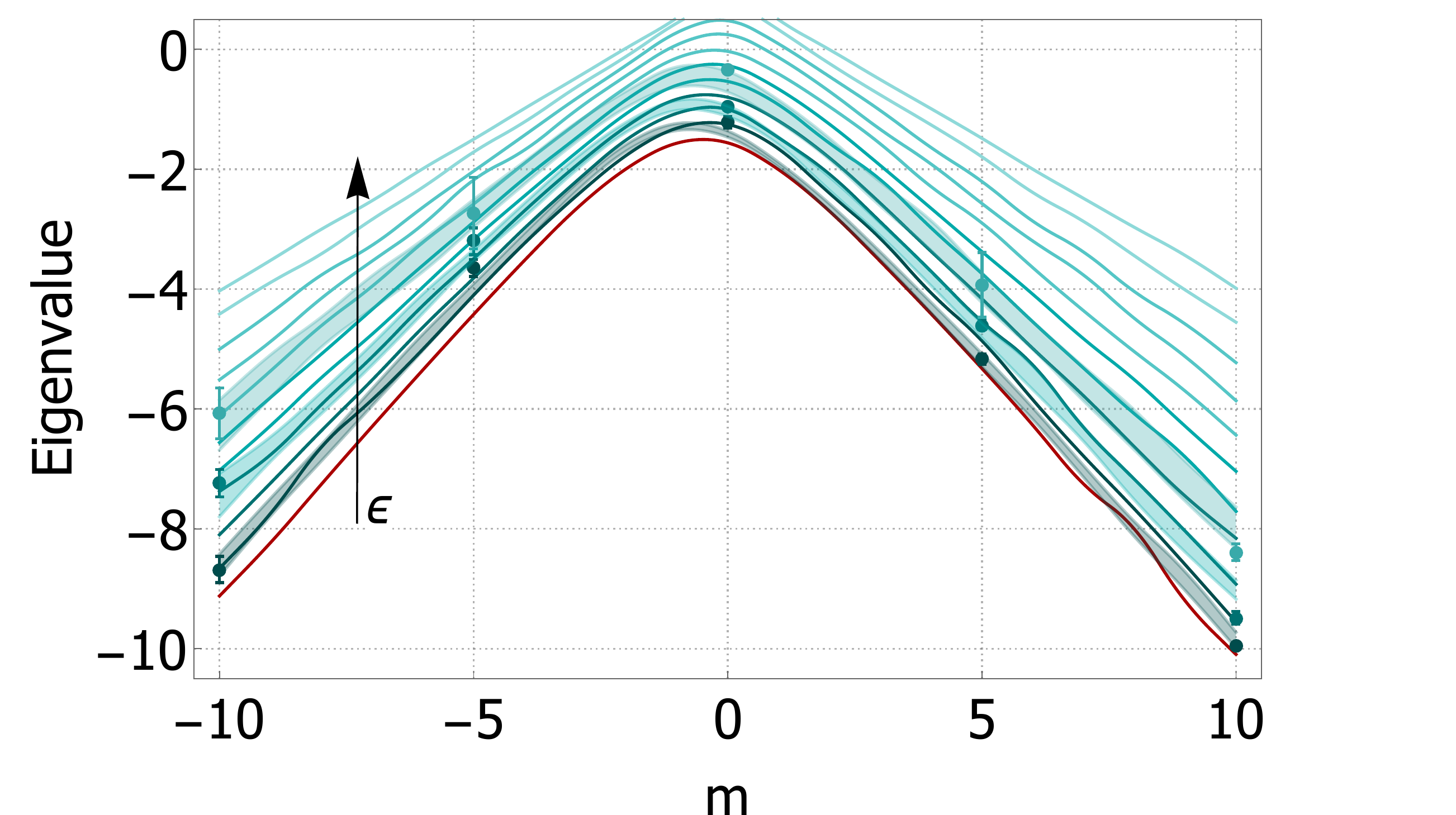}
        \label{fig:ev_1ch}
   }
   \subfloat[]
   {
        \includegraphics[width=0.42\textwidth]{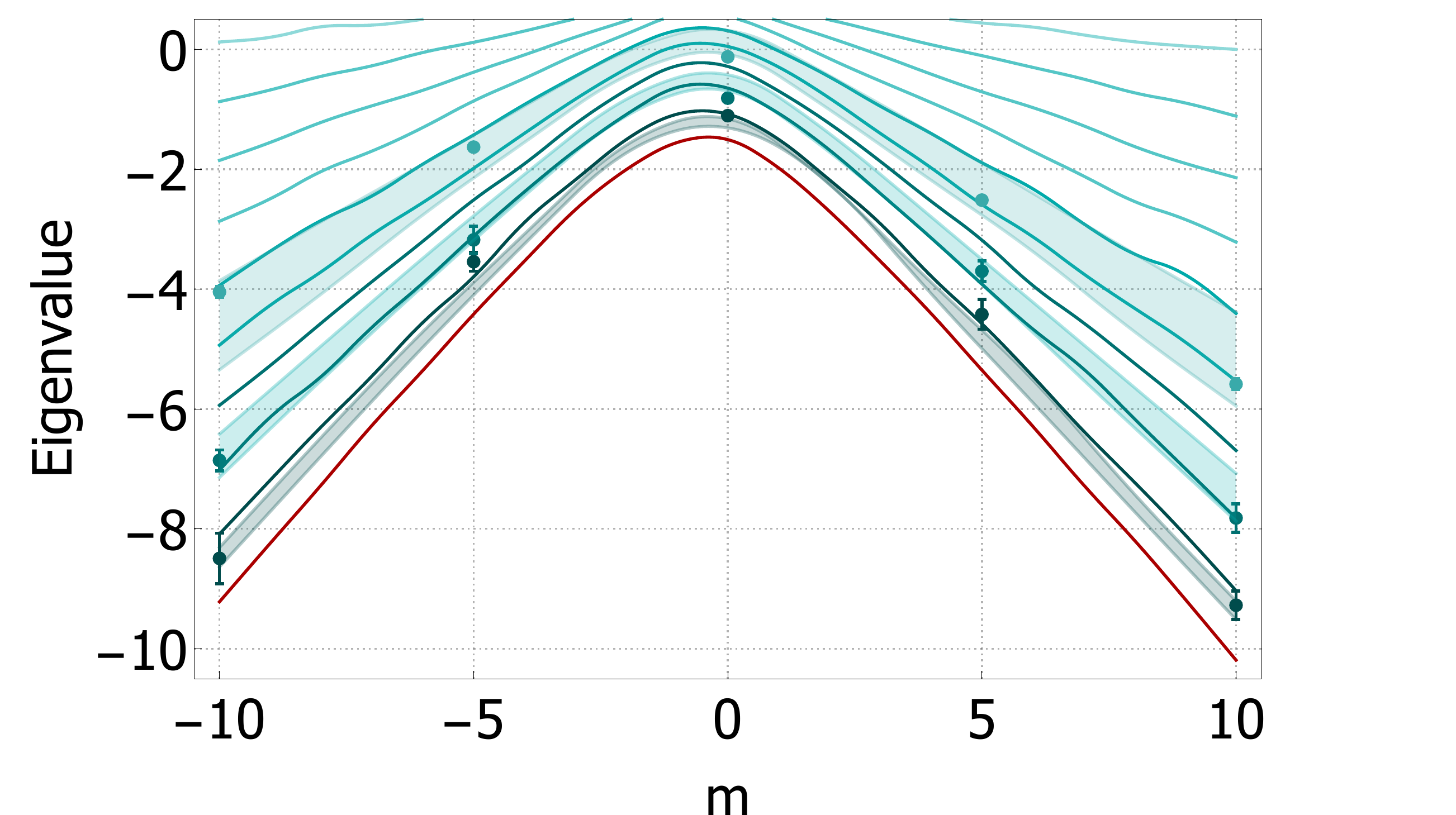}
        \label{fig:ev_2ch}
   }\\
    \subfloat[]
   {
        \includegraphics[width=0.42\textwidth]{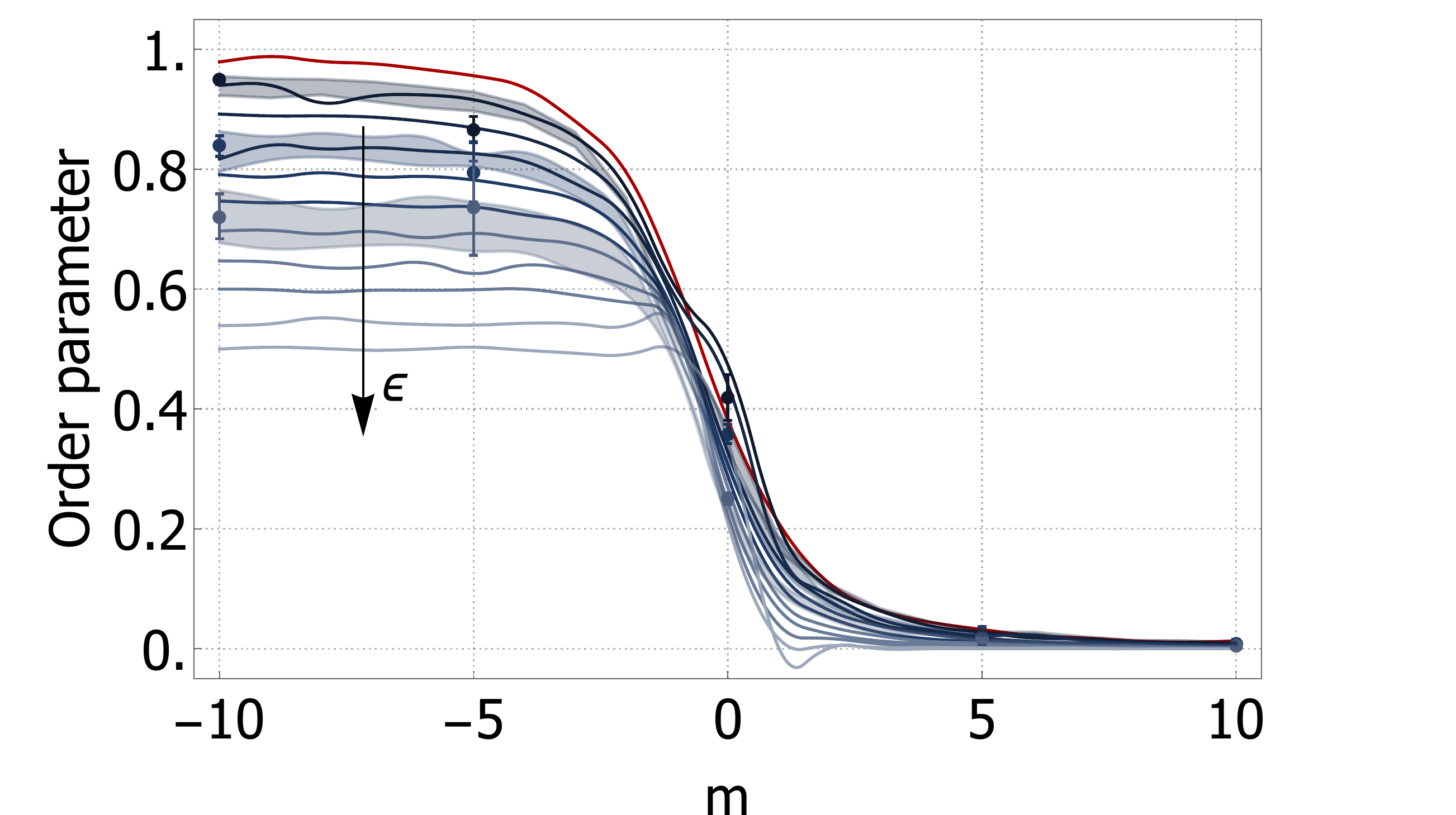}
        \label{fig:op_1ch}
   }
   \subfloat[]
   {
        \includegraphics[width=0.42\textwidth]{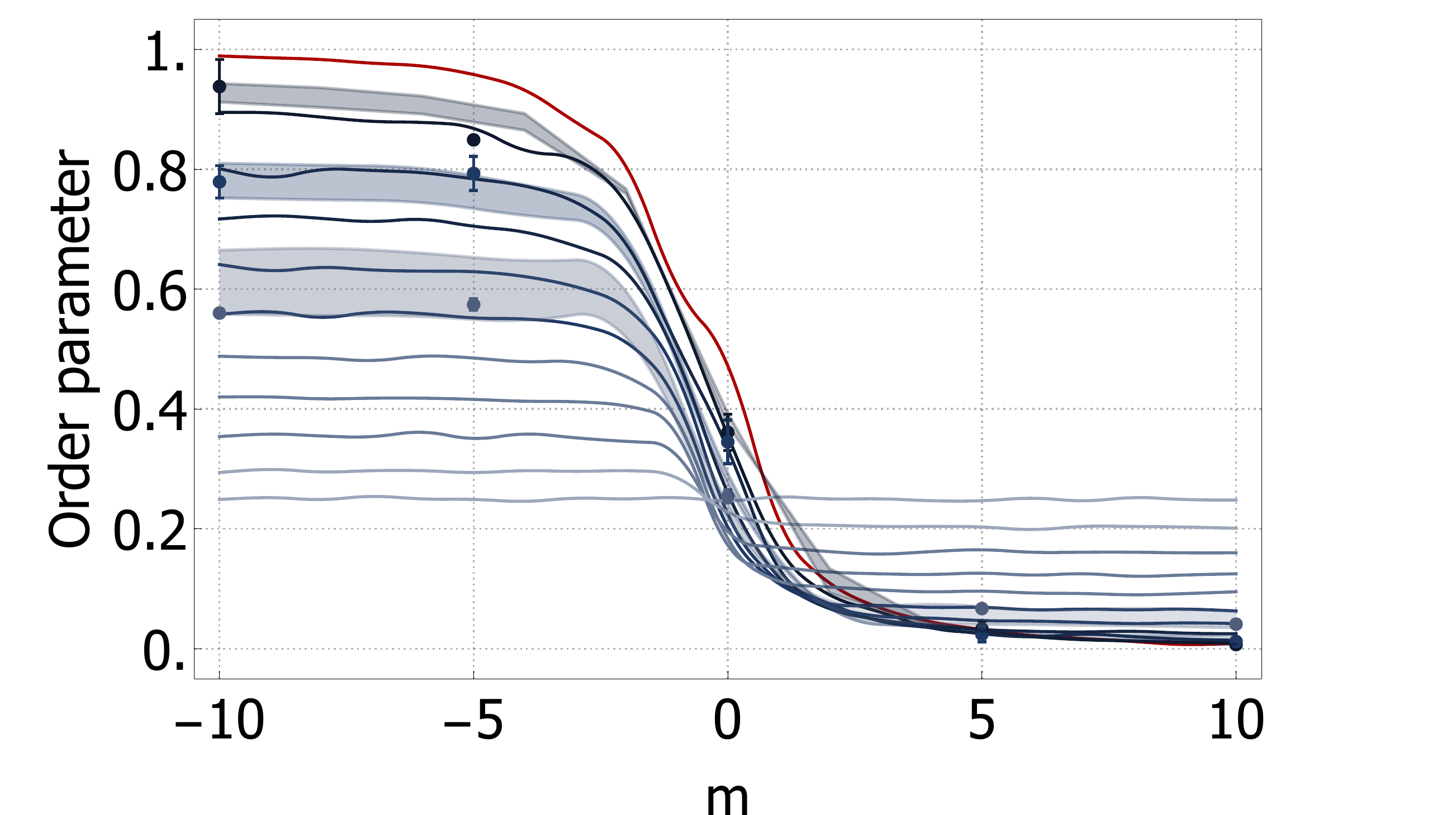}
        \label{fig:op_2ch}
   }
\caption{Noise simulations for dephasing of one qubit (a), (c) and both qubits (b), (d). Figs.~(a) and~(b) show the minimal eigenvalue dependence on $m$ and Figs.~(c) and~(d)---the dependence of the order parameter. Red lines  correspond to noiseless simulations, the color blur corresponds to the increase of noise strength $\epsilon$ from $0.1$ to $1$ in $0.1$ steps. Points---experimental results, solid areas---theoretical prediction for points with different noise level.}
\label{fig:noise_sim}
\end{figure*}

Compared to other types of quantum computers, photon circuits have low intrinsic noise levels. This means that we can add noise to the system in a controlled manner and get the dependencies of the parameters of interest on the noise level. We took advantage of this to evaluate the effect of noise on the phase transition that we observed without the noise. We expect that as the degree of dephasing increases, the phase transition will blur until it disappears completely. This will allow us to estimate the acceptable noise level in the system implementing VQE to identify quantum phase transitions.

The origin of the noise model used is connected with our experimental implementation. We artificially introduce noise to the system with liquid crystal variable retarders (LCVR) that are placed directly before Wollaston prisms adding noise to measurement. Placing them in the ansatz preparation part would require taking into account fiber transformations and would allow the algorithm to compensate the noise. LCVRs allow us to change the phase of the specific polarization component of the light field. If the phase shift~$\delta$ varies during the data acquisition time, then this leads to effective decoherence of the system state. The noise channel~$\mathcal E(\rho)$ is thus the transformation~\eqref{eq:Waveplate} averaged over~$\delta$ taken from some interval (depending on the noise strength). The explicit action of the noise channel is
\begin{gather}
\rho' = \mathcal E(\rho) = \sum_{j=1}^{2} E_j \rho E_j^{\dagger},\\
E_i = V(\theta) D_i(\delta) V^\dagger(\theta),\nonumber\\
D_1(\delta) = \sqrt{\frac{2-\epsilon}{2}} \begin{pmatrix} e^{i\delta}& 0 \\ 0 & 1 \end{pmatrix}, \quad
D_2(\delta) = \sqrt{\frac{\epsilon}{2}} \begin{pmatrix} e^{i\delta} & 0 \\ 0 & -1 \end{pmatrix},\nonumber
\end{gather}
where $E_j$ are the Krauss operators, $\theta$ is a LCVR axis angle, and $\delta$ is a mean retardance. Noise strength is controlled by the parameter~$\epsilon$, $0 \leq \epsilon \leq 1$. We set $\theta = \pi/4$ and $\delta = 2 \pi$ in our experiment.

Our experimental setup allows to explore the effect of this noisy channel on one qubit or simultaneously on both. Primarily, we simulated these two cases for different noise levels $\epsilon$ ranging from $0$ to $1$ with a $0.1$ step to obtain the eigenvalues and the values of the order parameter versus $m$. As expected, the presence of noise in the system prevents the algorithm from converging to the exact eigenvalue, and noise escalation leads to convergence deterioration (Fig.~\ref{fig:noise_sim}).

Finding appropriate eigenvalue becomes challenging for the case of simultaneous dephasing in both channels, and full dephasing ($\epsilon = 1$) leads to degeneracy---the algorithm converges to $1$ for any $m$. The phase transition in the order parameter blurs with increasing noise and disappears for $\epsilon = 1$. Full dephasing makes the order parameter constant and equal to $\braket{O} = 1/4$ for any $m$. In the case of a single noise channel the phase transition remains visible even with $\epsilon = 1$, while the maximum value of order parameter is halved.

 
Quantum phase transitions as metal-insulator transition and transition between quantum Hall liquid states, can be predicted and inquired by quantum algorithms. As we experimentally demonstrated, noise does not impede the detection of the phase transition point in a large range of noise levels. Only completely dephasing channels acting on both qubits prevent finding it in our model. This result demonstrates the noise-tolerance of VQE not only from speed and quality of convergence perspective but also from a practical point of view of determining the parameters of the Hamiltonian corresponding to a quantum phase transition.

We observe slow VQE convergence near the phase transition point and connect this behavior with the Hamiltonian ground state's closeness to the two-qubit singlet state. It seems to be a common effect for a combination of sufficiently general ansatz circuits and Hamiltonians, where the ground state exhibits additional symmetry. This hypothesis should be verified in future research. Possible approaches to circumvent poor convergence may include QAOA~\cite{farhi2014quantum}, because it uses specific ansatz adjusted for the target Hamiltonian.

Scalability is a major challenge for all modern quantum computing platforms, with the photonic one not being an exception. The experimental approach taken here may be relatively straightforward scaled up to 6-10 photons, and experiments on such scales are feasible \cite{wang2016experimental}. When the system is scaled up to larger number of photons, polarization encoding and free-space implementation used here is probably not the best option, and one should aim at integrated photonic circuits. Here we may note, that large fully programmable circuits are available and they can be used to realize parametrized transformations for variational algorithms~\cite{carolan2020variational}. The exact forms of optimal variational ansatze for such encoding are not yet known, and are an area of future research. However, a standard practice of using dual-rail encoded qubits allows one to realize variational algorithms as described here at an expense of finite probability of multi-qubit operations, which may be brought close to unity with an addition of extra photons~\cite{Milburn_RevModPhys}

The major challenge on the experimental side for large scale experiments will be photon loss in the circuit, which dramatically reduces the count rate for multi-photon events. So on short time scales one should aim for higher brightness single photon sources and low-loss integrated optics. On a longer timescale a fully integrated modular architecture for photonic computing may be developed, which has intrinsic loss tolerance, since the path length of each photon becomes independent of the circuit size due to the modular structure of the processor~\cite{bartolucci2021fusion}.

\begin{acknowledgments}


The Skoltech team acknowledges support from the research project, {\it Leading Research Center on Quantum Computing} (agreement No.~014/20). The MSU team acknowledges financial support from the Russian Foundation for Basic Research (RFBR Project No. 19-32-80043 and RFBR Project No. 19-52-80034) and support under the Russian National Technological Initiative via MSU Quantum Technology Centre.
{\it \bf Competing interests:} The authors declare no competing interests.
{\it \bf Data and code availability:} The data that supports this study are available within the article. The code for generating the data will be made available on GitHub after this paper is published.

\end{acknowledgments}
\onecolumngrid
\newpage
\section*{Appendix}
\appendix
\subfile{aipsamp_sup}

\bibliography{aipsamp}

\end{document}

%% file: aipsamp_sup.tex
\renewcommand{\thefigure}{S\arabic{figure}}
\renewcommand{\thesection}{S\arabic{section}}  
\renewcommand{\theequation}{S\arabic{equation}}  
\preprint{AIP/123-QED}

\title[Variational Simulation of Schwinger's Hamiltonian with Polarisation Qubits]{Supplementary material:\\Variational Simulation of Schwinger's Hamiltonian with Polarisation Qubits}

\author{O.\,V.~Borzenkova}\email{oksana.borzenkova@skoltech.ru}
\affiliation{Skolkovo Institute of Science and Technology, 3 Nobel Street, Moscow 121205, Russian Federation}
\author{G.\,I.~Struchalin}
\affiliation{Quantum Technology Centre and Faculty of Physics,  M.V.~Lomonosov Moscow State University,  1 Leninskie Gory Street, Moscow 119991, Russian Federation} 
\author{A.\,S.~Kardashin}
\affiliation{Skolkovo Institute of Science and Technology, 3 Nobel Street, Moscow 121205, Russian Federation}
\author{V.\,V.~Krasnikov}
\affiliation{Quantum Technology Centre and Faculty of Physics,  M.V.~Lomonosov Moscow State University,  1 Leninskie Gory Street, Moscow 119991, Russian Federation} 
\author{N.\,N.~Skryabin}
\affiliation{Quantum Technology Centre and Faculty of Physics,  M.V.~Lomonosov Moscow State University,  1 Leninskie Gory Street, Moscow 119991, Russian Federation} 
\author{S.\,S.~Straupe}
\affiliation{Quantum Technology Centre and Faculty of Physics,  M.V.~Lomonosov Moscow State University,  1 Leninskie Gory Street, Moscow 119991, Russian Federation} 
\author{S.\,P.~Kulik}\homepage{https://quantum.msu.ru}
\affiliation{Quantum Technology Centre and Faculty of Physics,  M.V.~Lomonosov Moscow State University,  1 Leninskie Gory Street, Moscow 119991, Russian Federation} 
\author{J.\,D.~Biamonte}\homepage{https://quantum.skoltech.ru}
\affiliation{Skolkovo Institute of Science and Technology, 3 Nobel Street, Moscow 121205, Russian Federation}

\date{\today}
\maketitle

\onecolumngrid

\section{Classical optimizer\label{sec:SPSA}}
The target function under minimization $E(\boldsymbol\theta)$ is the mean Hamiltonian value, but in the experiment, only random samples of $E(\boldsymbol\theta)$ obtained by repetitive measurements are available. So experimental VQE is a \emph{stochastic approximation} problem~\cite{Kushner_Book1997}. We use a simultaneous perturbation stochastic approximation (SPSA) algorithm~\cite{Spall_TAC1992} as a classical optimizer in our VQE implementation. It is useful for high-dimensional problems, where the gradient of the objective function is not directly available, because SPSA requires only two function evaluations per iteration for any number of parameters in the optimization problem.

Single SPSA iteration proceeds as follows:
\begin{enumerate}
    \item Generate a random vector $\boldsymbol\Delta$ with elements being $\pm 1$ with equal probability.
    \item Estimate a gradient $\boldsymbol g$:
    \begin{equation}
        \boldsymbol g = \frac{E(\boldsymbol\theta + b \boldsymbol\Delta) - E(\boldsymbol\theta - b \boldsymbol\Delta)}{2 b} \boldsymbol\Delta.
    \end{equation}
    \item Move to the new point~$\boldsymbol\theta'$:
    \begin{equation}
        \boldsymbol\theta' = \boldsymbol\theta - a \boldsymbol g.
    \end{equation}
\end{enumerate}
Scalar variables $a$ and $b$ are called \emph{meta parameters}. The parameter $a$ describes the iteration step and $b$ defines finite difference to calculate the gradient. They change with the number of iterations~$k$ according to schedule:
\begin{equation}
    a(k) = \frac{a_0-a_f}{k^{0.602}} + a_f,\quad
    b(k) = \frac{b_0-b_f}{k^{0.101}} + b_f.
\end{equation}

Usually final values $a_f$ and $b_f$ are set to zero to ensure convergence in the limit $k \to \infty$. However, we use nonzero $a_f$ and $b_f$ to track a slow drift of the experimentally prepared probe state $\ket{\psi(\boldsymbol\theta)}$ over time~\cite{Granichin_TAC2015}. The drift occurs mainly due to the instability of polarization transformation in optical fibers connecting the SPDC source and measurement part of the setup.

Moreover, we find out influence of mass parameter~$m$ on VQE convergence---closeness to phase transition makes it slowly. So we adjust meta parameters for each $m$ as
\begin{equation}
    a_{0,f}(m) = \frac{\bar a_{0,f}}{0.2 m + 1},\quad
    b_{0,f}(m) = \frac{\bar b_{0,f}}{0.2 m + 1}
\end{equation}
In our simulations and the experiment we used $\bar b_0 = 0.1$, $\bar b_f = 0.002$ and tried different $\bar a_0$ and $\bar a_f$ to find trade-off between the number of iterations and accuracy. For $\bar a_0 = 0.01$ and $\bar a_f = 0.003$ convergence is slow, especially in the experiment. After different simulations we chose $\bar a_0 = 0.05$ and $\bar a_f = 0.005$.

\section{Convergence\label{sec:Convergence}}
\begin{figure}[!htb]
    \centering
    \subfloat[$m=-8$]{
    \includegraphics[width=0.4\textwidth]{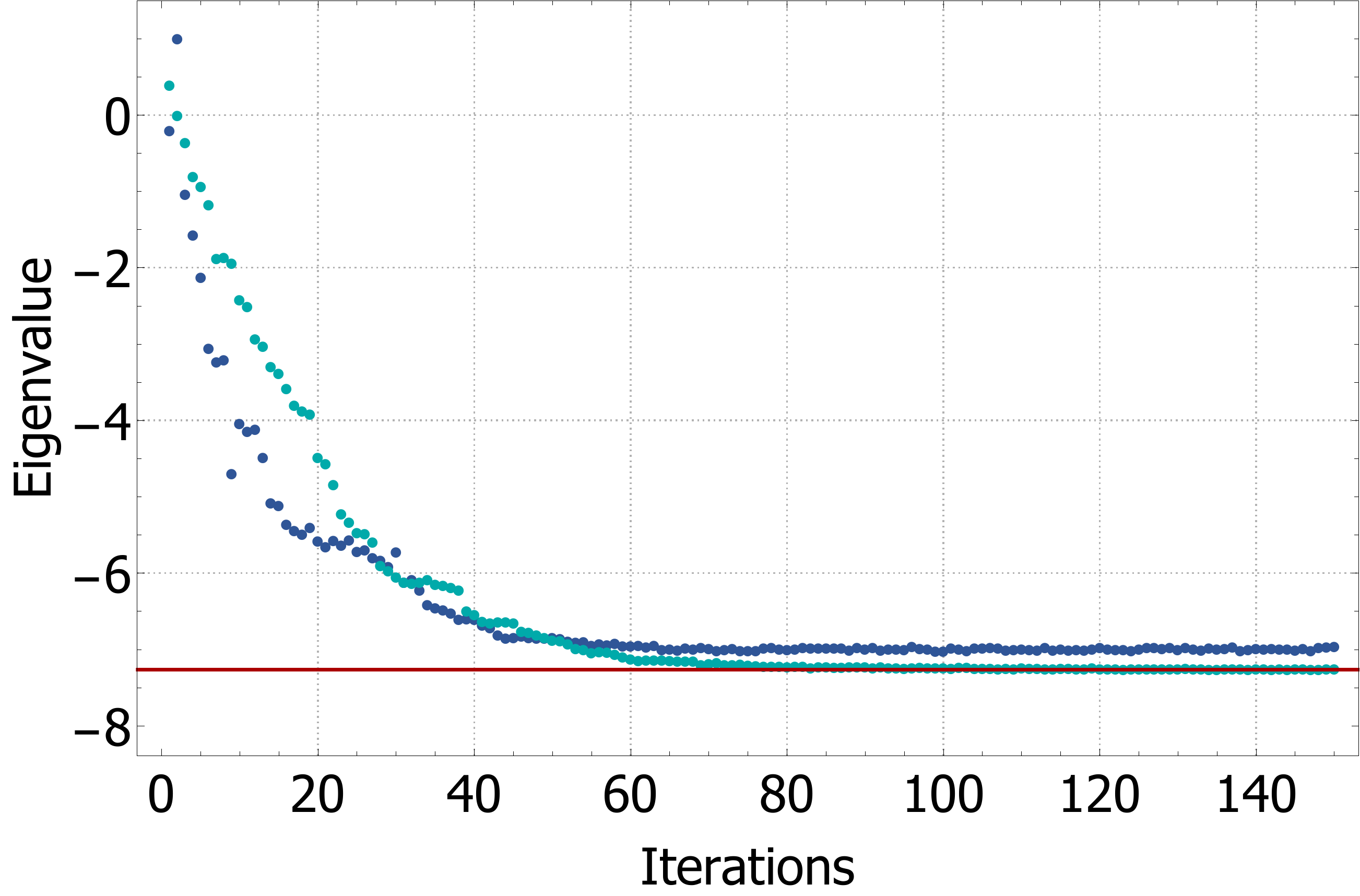}
    \label{fig:convergence_m-8}}
    \subfloat[$m=2$]{
    \includegraphics[width=0.412\textwidth]{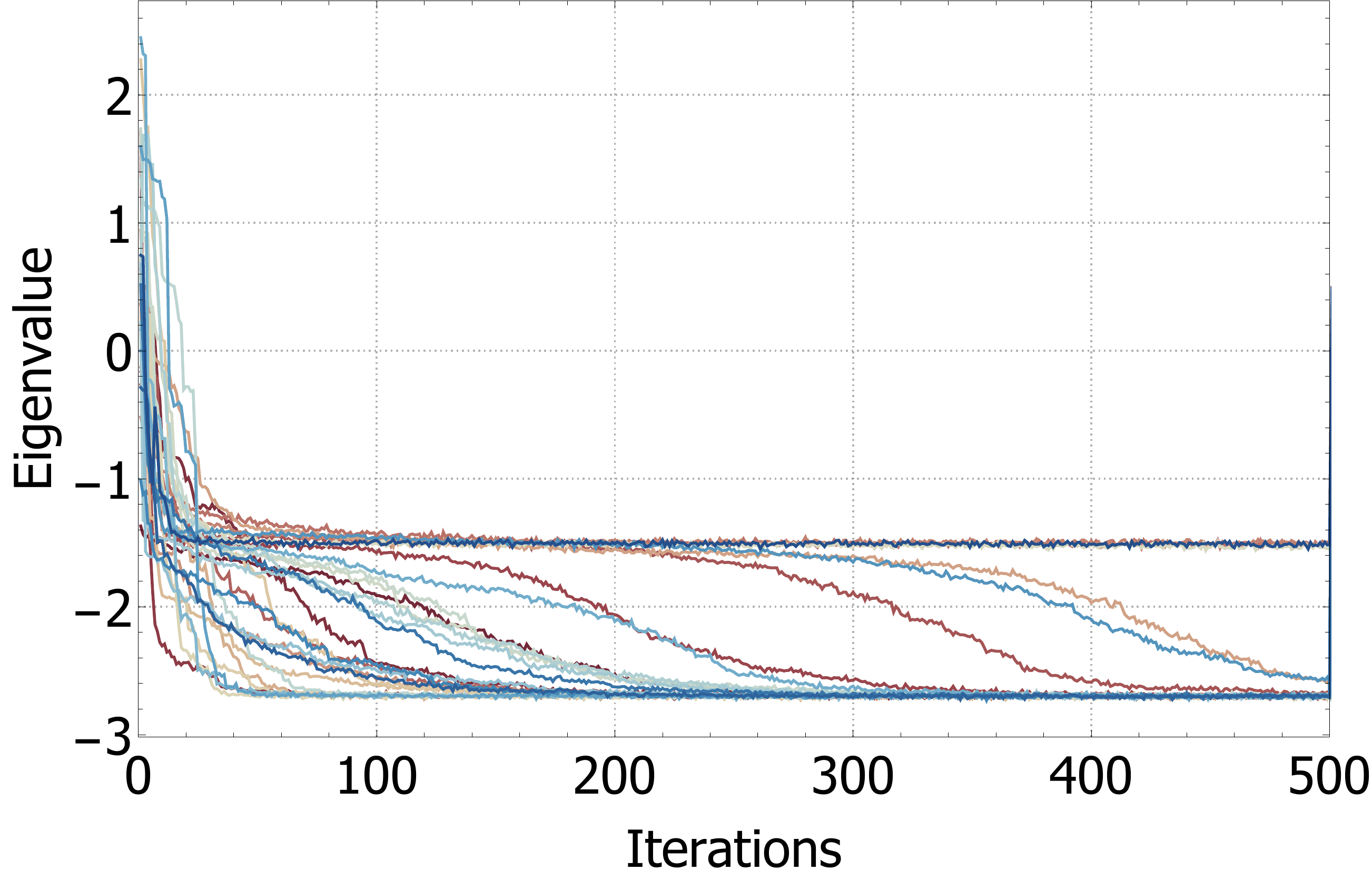}
    \label{fig:convergence_m_2}}
    \caption{(a) VQE convergence for $m=-8$: cyan points---median of 30 simulation runs, blue points---median of 3 experimental runs, red line---exact energy value. (b) The algorithm convergence for $m=2$ for all 30 runs.}
    \label{fig:convergence}
\end{figure}

Figure \ref{fig:convergence}(a) demonstrates VQE convergence in the experiment (blue points) and in simulations (cyan points) for $m = -8$, which is far from the transition point. Both datasets represent the median value of different trials: 3 trials for the experiment and 30---for simulations. The number of iterations for the experiment and simulations in Fig.~3 varies: it takes 150 iterations in the experiment and 500 in simulations to converge. However, the final experimental value would be the same for 500 iterations. Furthermore, such a large number of iterations for simulations in Fig.~3 allows examination of how attractor (plateau) works. You can see the probability distribution transformation for different $m$: from unimodal (Gaussian-like) form, when $m$ is far from the transition point, to two-peak distribution in the center. Convergence plots in Fig.~\ref{fig:convergence}(b) demonstrate this situation with the number of iterations for $m=2$: some trials fall into the local minimum, and others reach the global one. Interestingly, after being stuck in the local minimum for a while, the SPSA algorithm can still converge to the global one. In our opinion, the stochastic manner of the algorithm may be the reason for such behavior.

\section{Initial point statistics\label{sec:Statistics}}

We carried out numerical simulations of the VQE algorithm for $m = 0, -1/2, 1, 10$ to investigate how the choice of an initial point $\boldsymbol\theta_0$ affects convergence and explore the set of obtained solutions. Recall that the Schwinger Hamiltonian $  H_2$ undergo phase transition of the order parameter at $m = -1/2$, so points $m = 0$ and $m = 1$ are nearby and symmetric w.\,r.\,t. phase transition and $m = 10$ is an example of a distant point. To collect statistics, we execute the VQE algorithm $10^5$ times for each $m$ starting from freshly generated random initial points~$\boldsymbol\theta_0$. The points are distributed uniformly in a six-dimensional hypercube with the side length equal to $\pi$, which coincides with the period of the target function $E(\boldsymbol\theta)$.

\begin{figure}[!htb]
    \centering
    \includegraphics[width=0.4\textwidth]{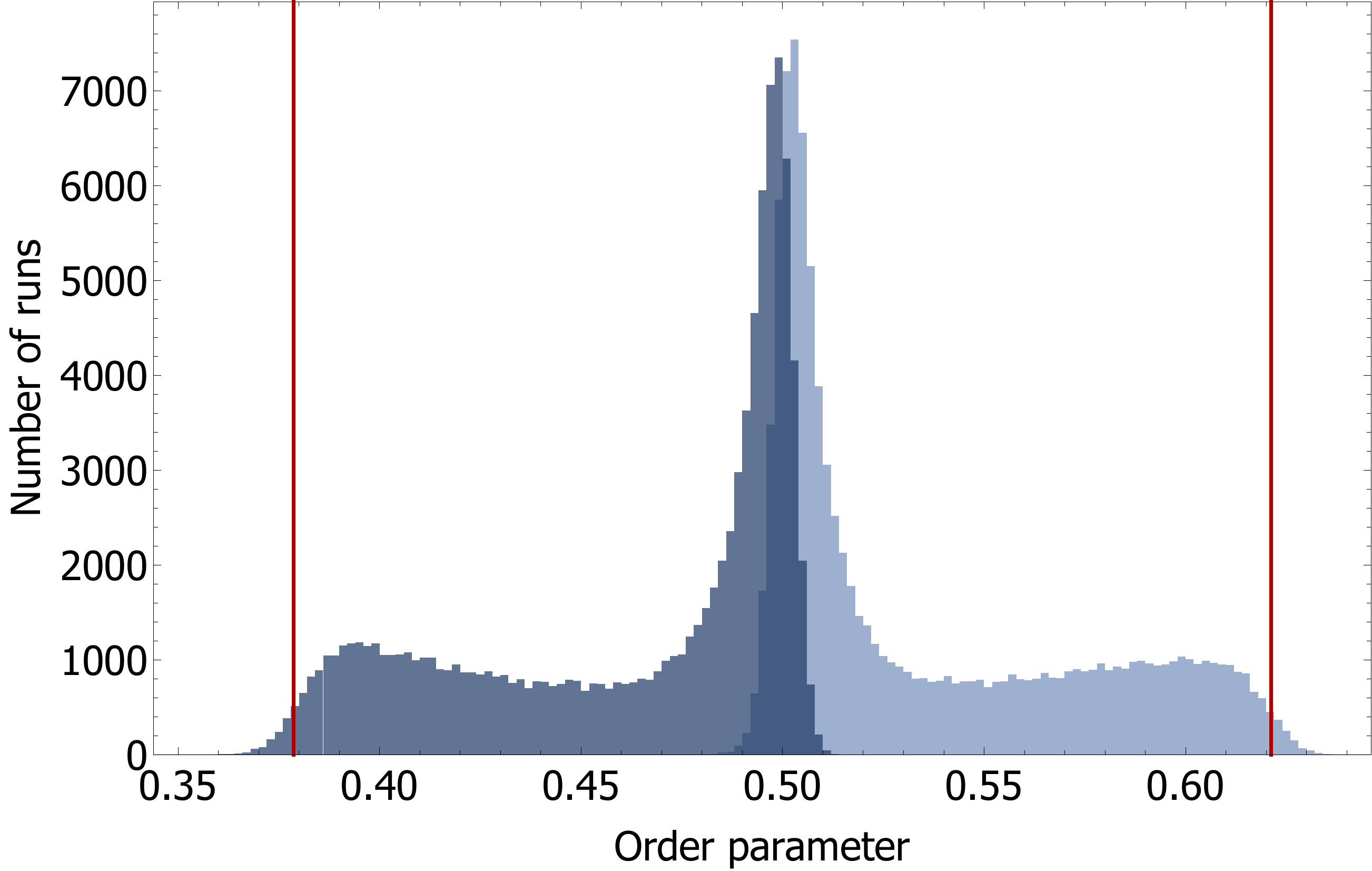}
    \caption{Order parameter histograms for $m = 0$ (darker, left) and $m = -1$ (lighter, right) obtained by results of $10^5$ VQE runs. Red vertical lines present analytical solutions. The histograms are nearly a reflection of each other around $\braket{  O} = 1/2$.}
    \label{fig:hist_op_m0_m-1}
\end{figure}

Each VQE run results in the final point~$\boldsymbol\theta$, the energy level $E(\boldsymbol\theta)$ (eigenvalue), and the order parameter $\braket{  O}$. Fig.~\ref{fig:hist_op_m0_m-1} shows histograms of~$\braket{  O}$ for $m = 0$ and $m = 1$. As one can see, there is a sharp peak near a wrong value $\braket{  O} = 1/2$ for both histograms and obtuse peaks approaching true solutions $\braket{  O} \approx 0.38$ and $\braket{  O} \approx 0.62$ for $m = 0$ and $m = -1$, respectively. As it was said in the main text, $\braket{  O} = 1/2$ corresponds to the plateau in landscape of the target function, which acts as an attractor for the optimizer.

\begin{figure}[!htb]
    \centering
    \subfloat[$m = 10$]
    {
        \includegraphics[width=0.4\textwidth]{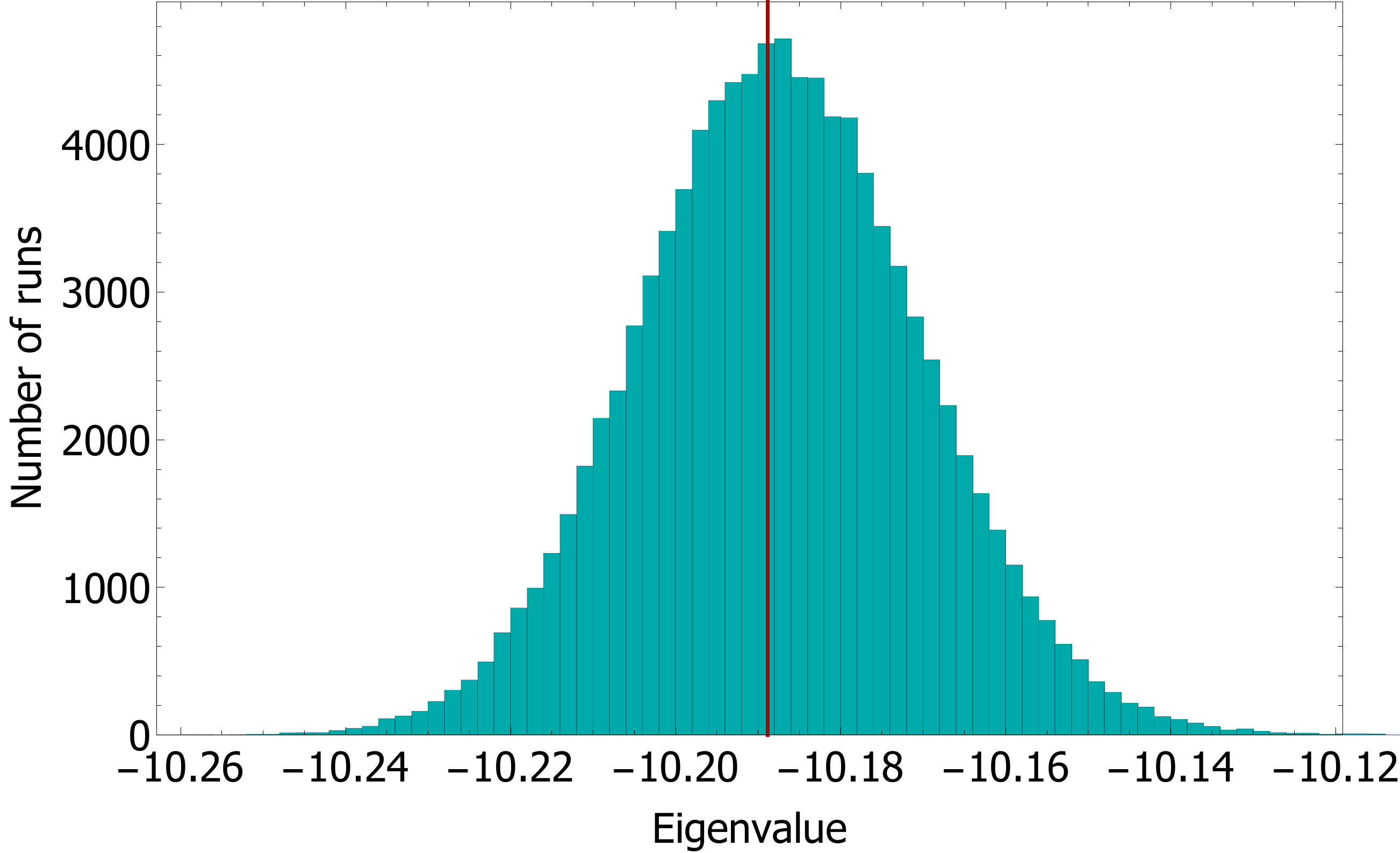}
        \label{fig:hist_m10}
    }\\
    \subfloat[$m = 1$]
    {
        \includegraphics[width=0.4\textwidth]{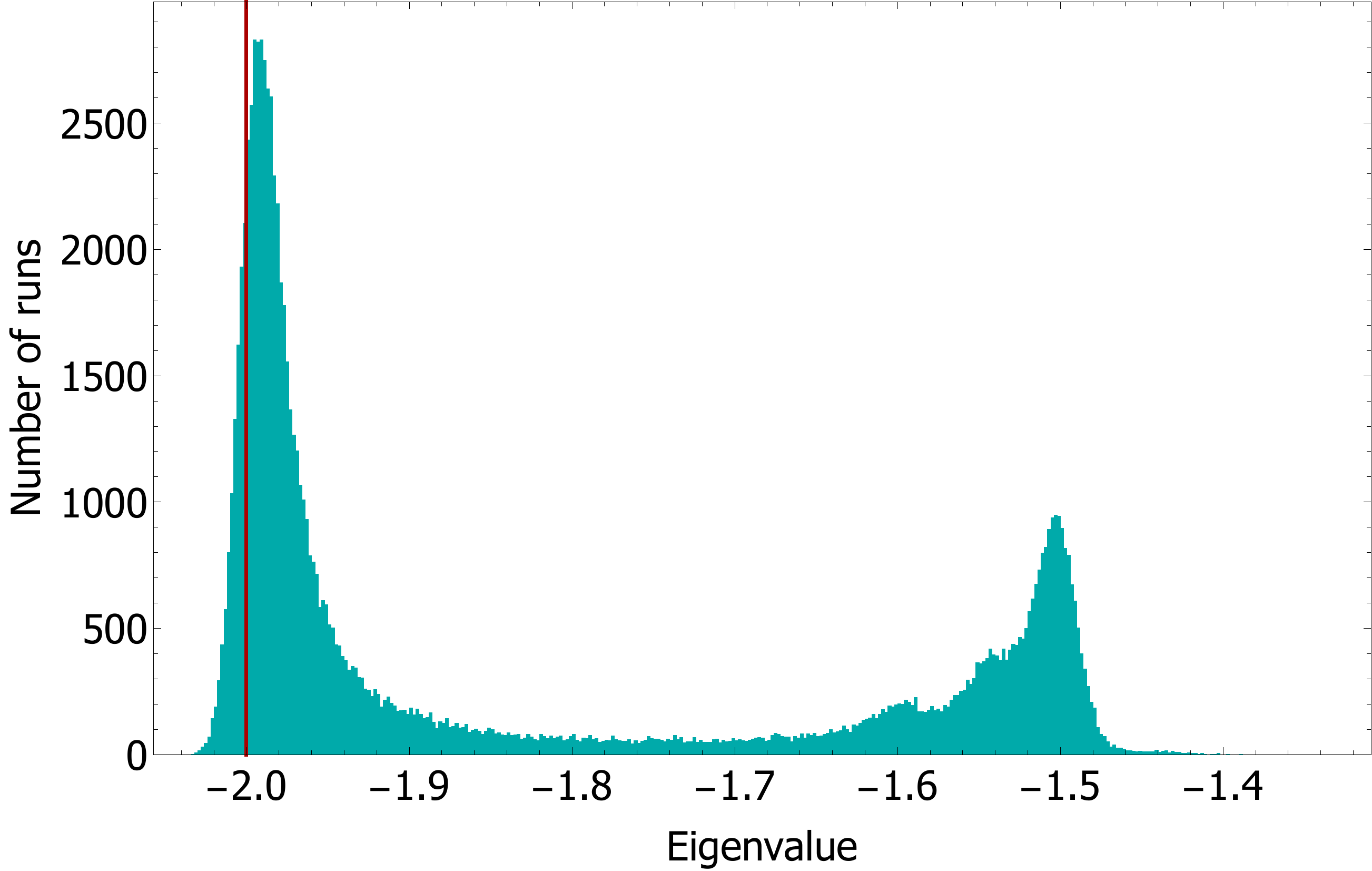}
        \label{fig:hist_m1}
    }\\
    \subfloat[$m = 0$]
    {
        \includegraphics[width=0.4\textwidth]{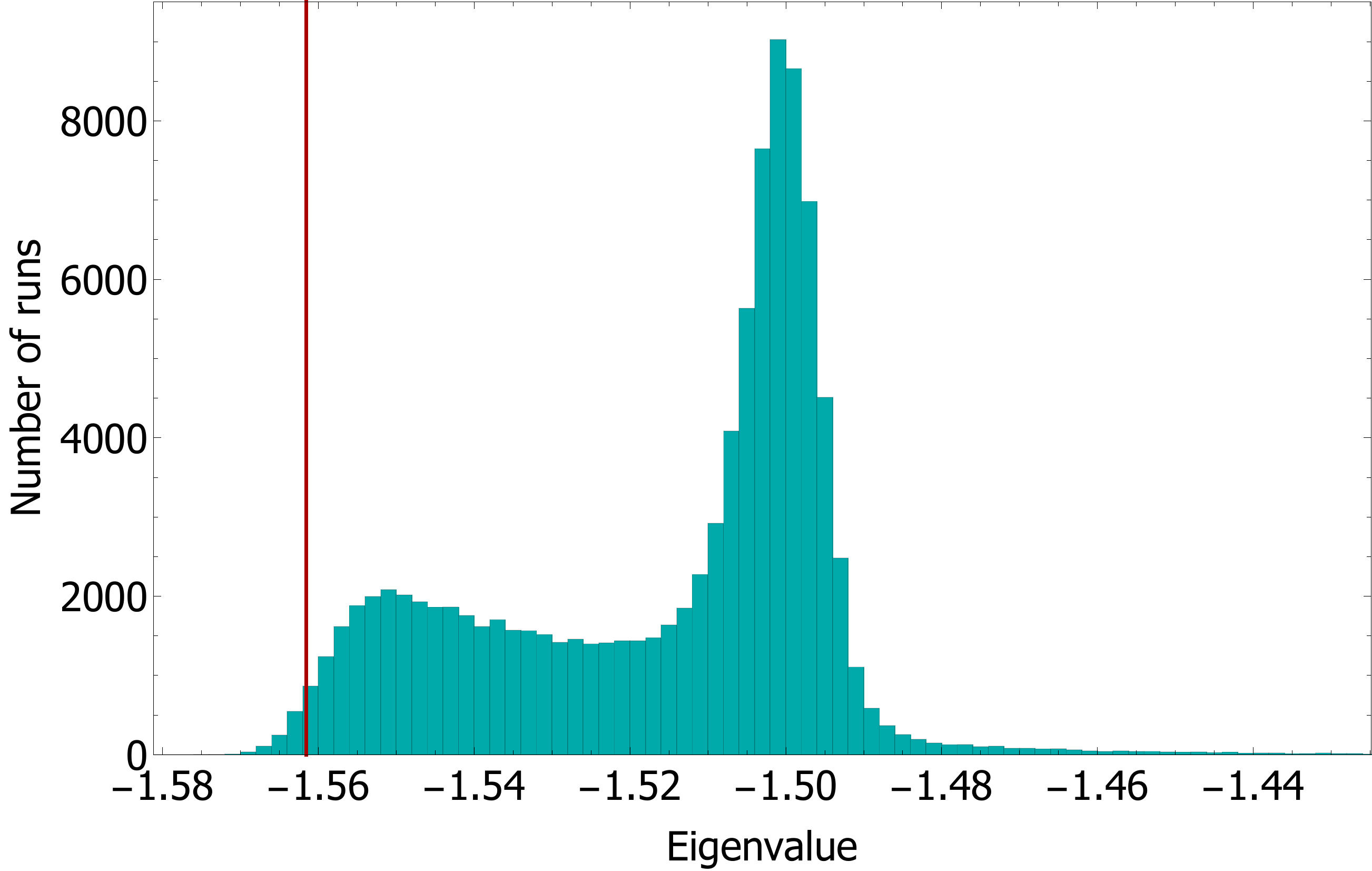}
        \label{fig:hist_op_m0}
    }
\caption{Histograms of eigenvalues that are found during $10^5$ VQE runs for $m = 10$ (a), $m = 1$ (b), and $m = 0$ (c). Red vertical lines show analytical solutions.}
\label{fig:statistics_hist}
\end{figure}

Fig.~\ref{fig:statistics_hist} illustrates evolution of found-eigenvalue distribution for different $m$. As expected, the histogram for $m = 10$ is unimoal and centered around the exact eigenvalue for the corresponding Hamiltonian. When~$m$ approaches phase transition point $m = -1/2$, the second peak emerges around $E = -3/2$, which is precisely the Hamiltonian eigenvalue for $m = -1/2$. This erroneous peak is small for $m = 1$, but it becomes even higher than the true one for $m = -1$. Closeness to phase transition point changes convergence statistics dramatically---less than $1\%$ of simulations reveal proper values for $m = -1$.

We tried to clarify the structure of obtained solutions in the space of tuned parameters~$\boldsymbol\theta$. First, we bring all found points~$\boldsymbol\theta$ to a hypercube that corresponds to the target function period. After that, for graphic purposes, we decreased the dimensionality of obtained solutions~$\boldsymbol\theta$ from six to three using principal component analysis (PCA). PCA finds a lower-dimensional hyperplane in the original space, which has a minimum average squared distance from the points to the hyperplane. Then points are projected to the approximating hyperplane. PCA helps to keep the real structure of the initial space and find any clusters of points with the same values.

\begin{figure*}[htpb]
    \centering
    \subfloat[Top, $m = -0.5$.]
    {
        \includegraphics[width=0.4\textwidth]{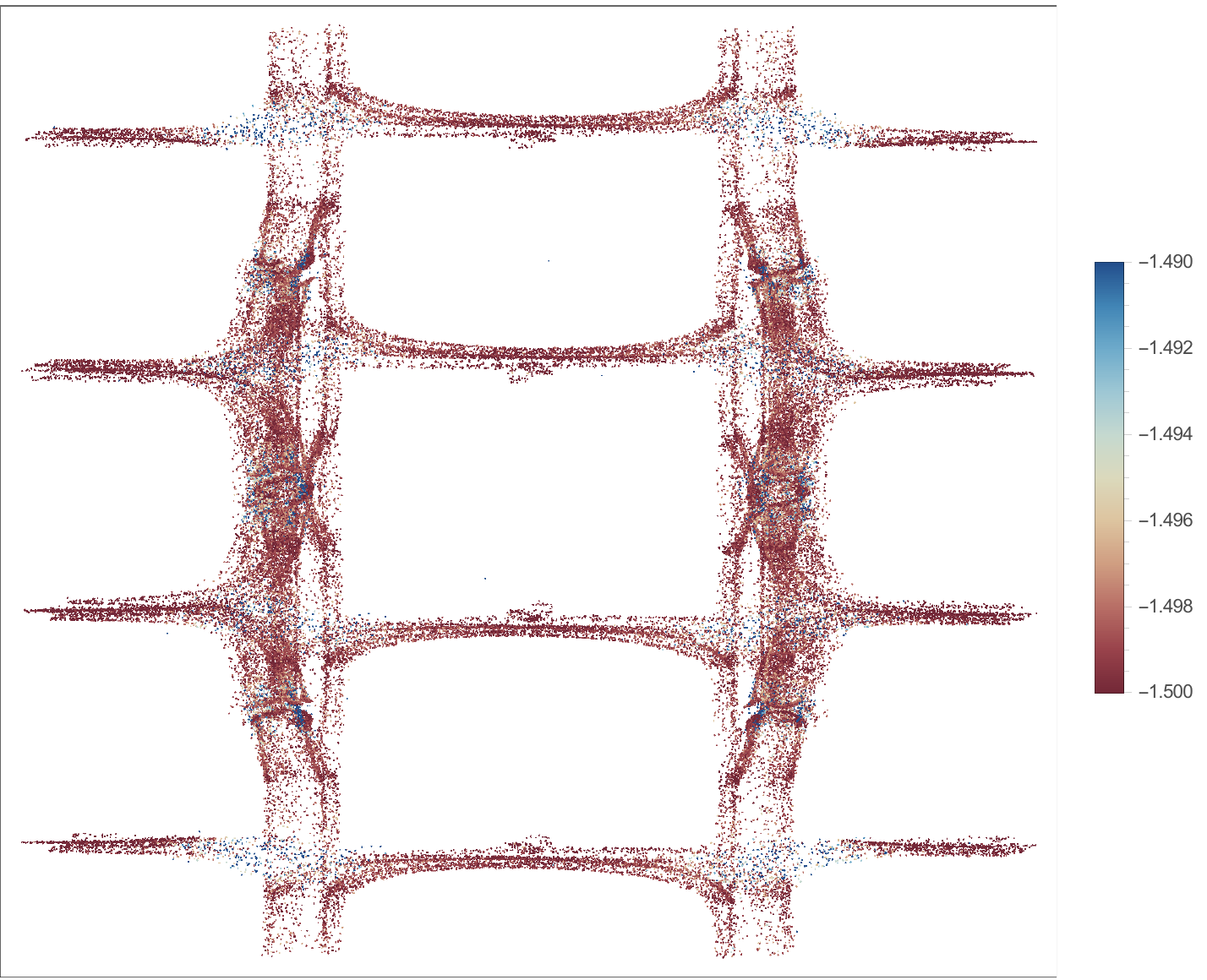}
        \label{fig:top-0.5}
    }
    \subfloat[Side, $m = -0.5$.]
    {
        \includegraphics[width=0.4\textwidth]{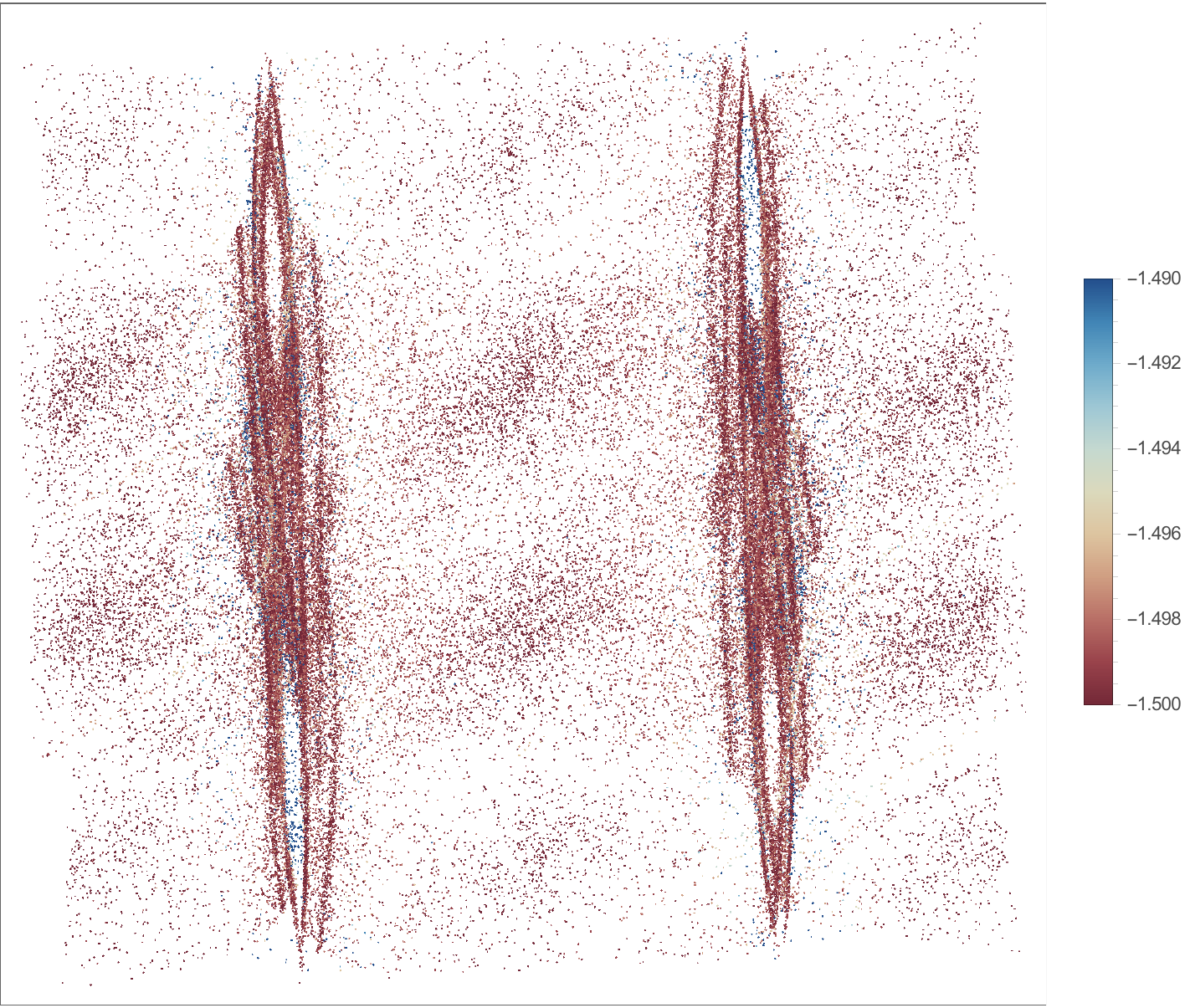}
        \label{fig:side-0.5}
    }\\
    \subfloat[Top, $m = 0$.]
    {
        \includegraphics[width=0.4\textwidth]{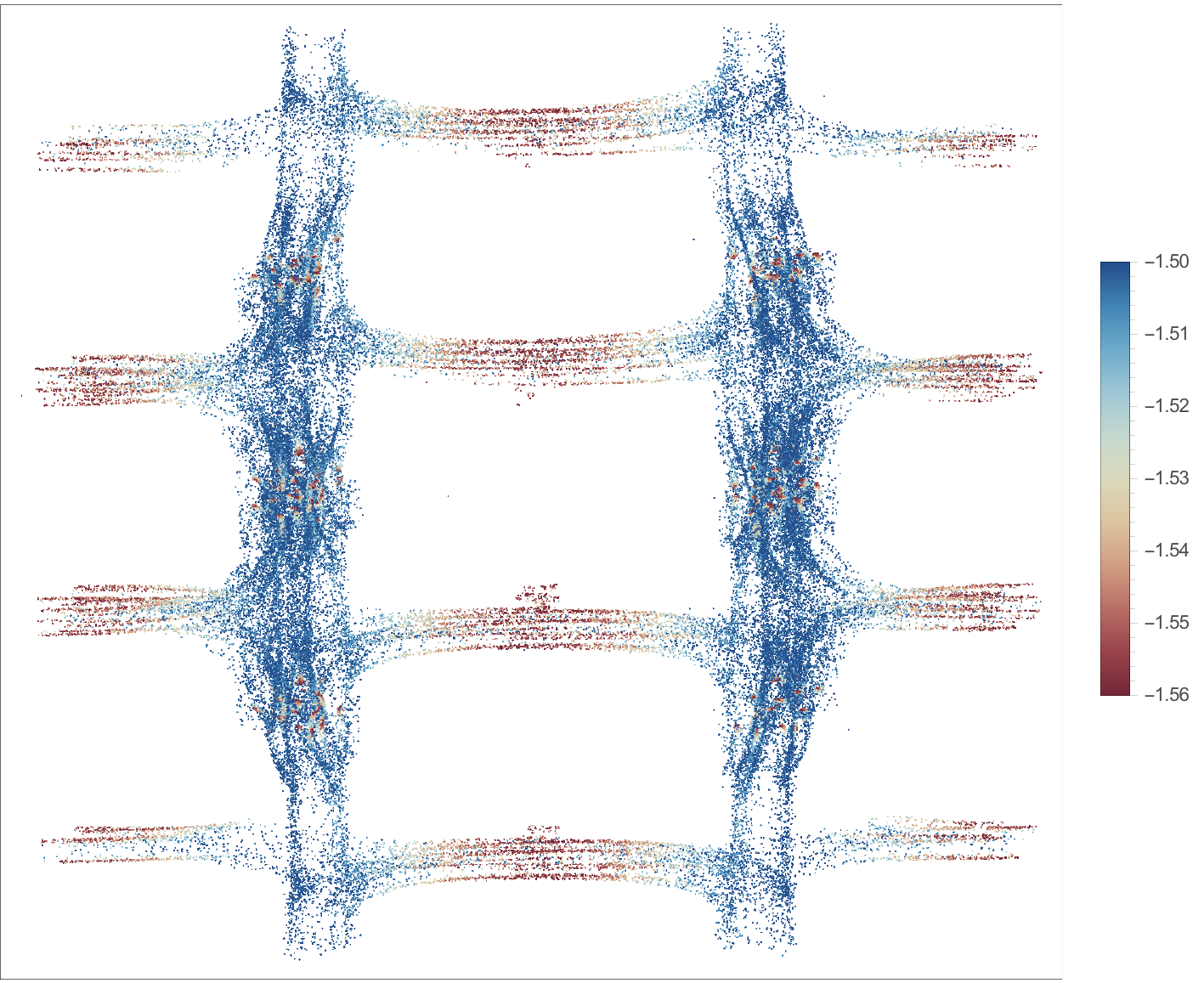}
        \label{fig:top0}
    }
    \subfloat[Side, $m = 0$.]
    {
        \includegraphics[width=0.4\textwidth]{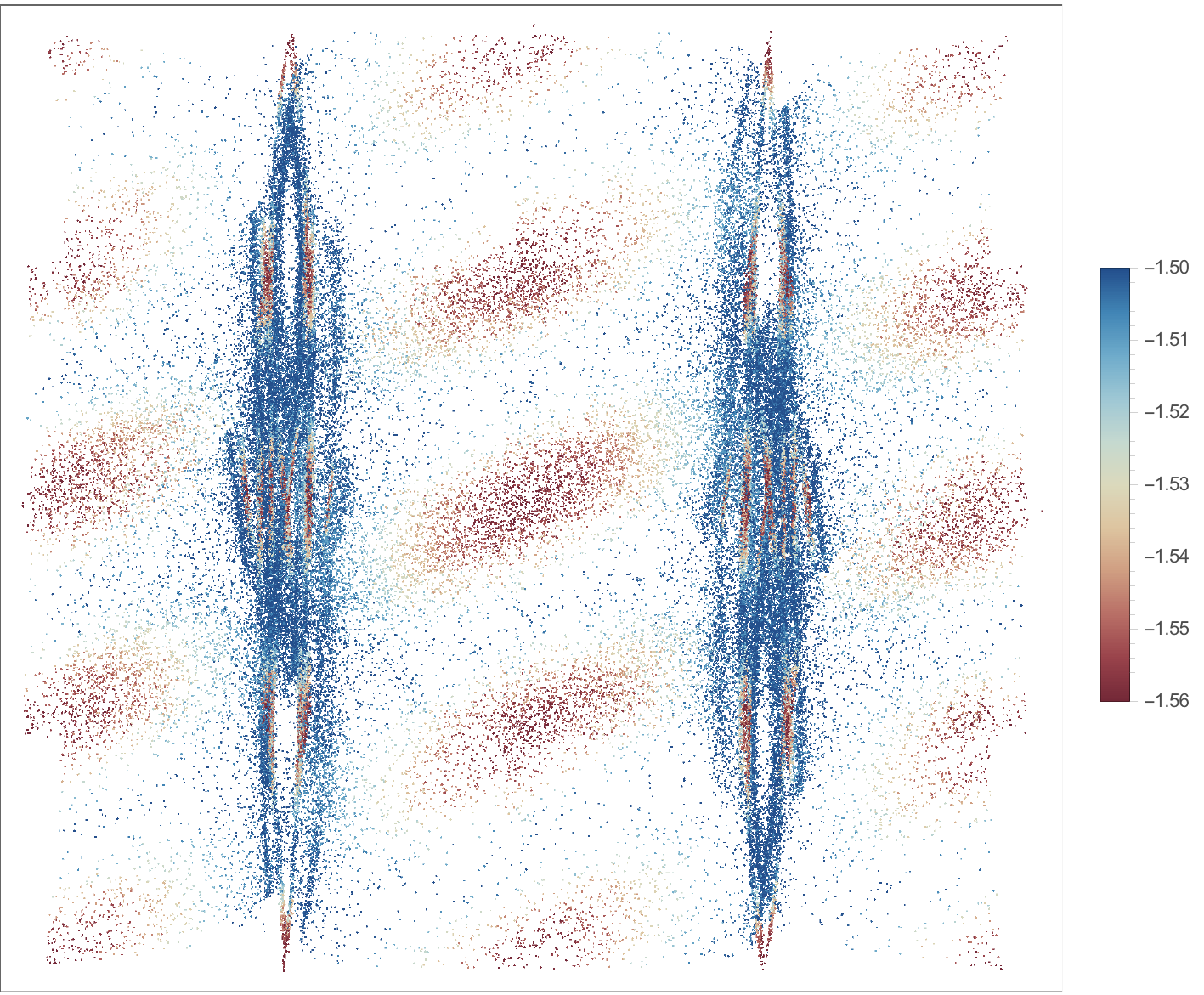}
        \label{fig:side0}
    }\\
    \subfloat[Top, $m = 1$.]
    {
        \includegraphics[width=0.4\textwidth]{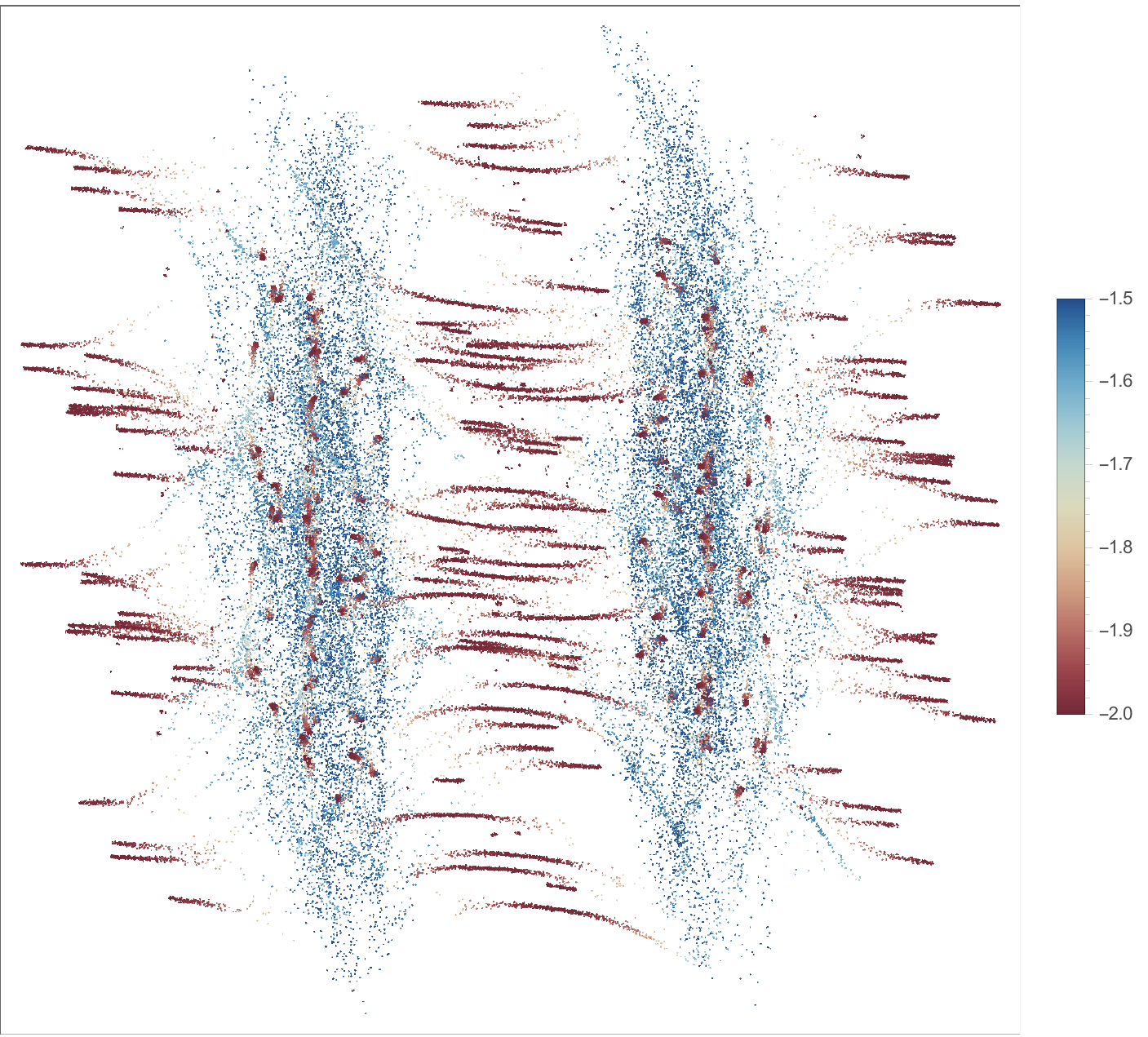}
        \label{fig:top1}
    }
    \subfloat[Side, $m = 1$.]
    {
        \includegraphics[width=0.4\textwidth]{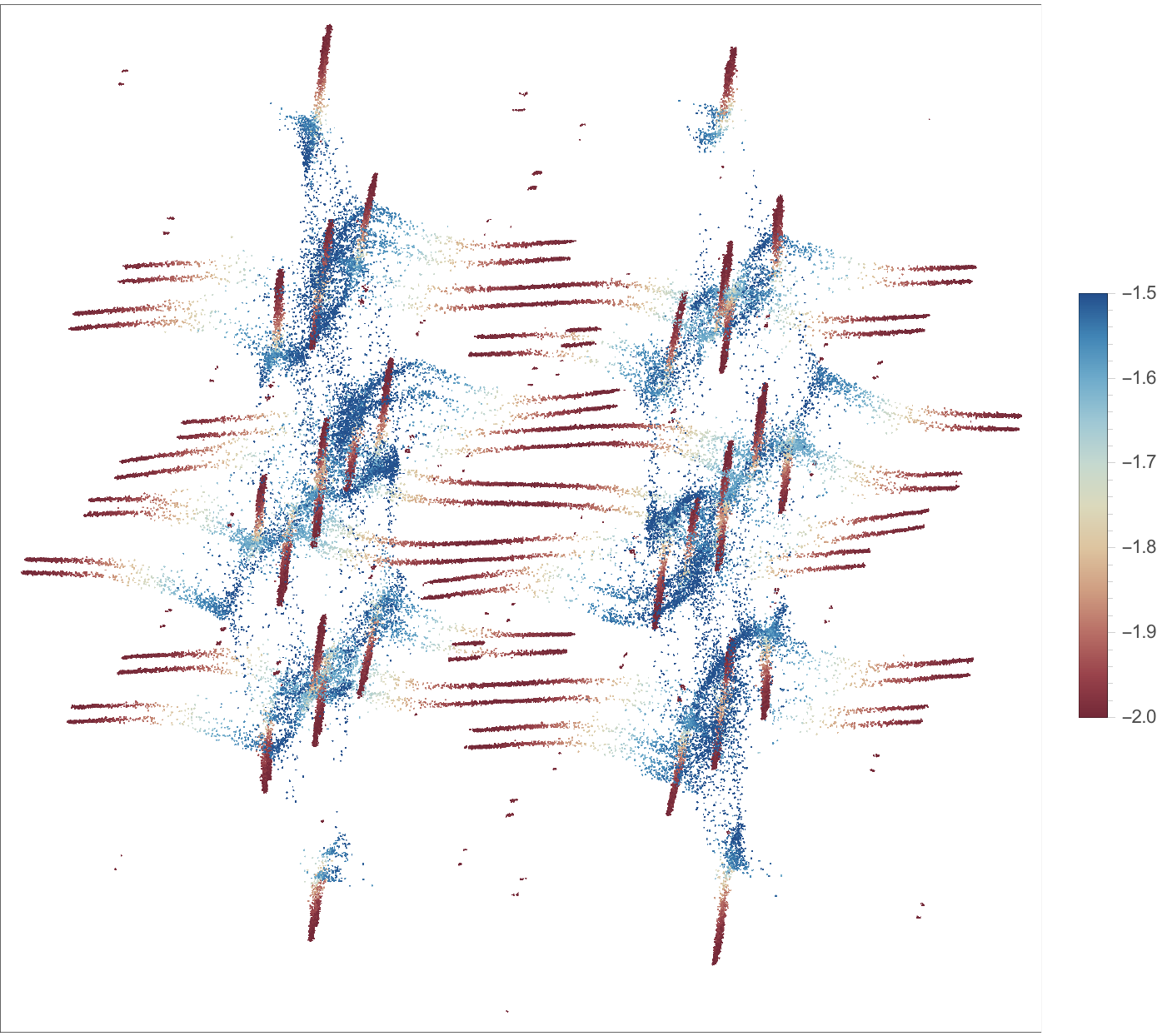}
        \label{fig:side1}
    }
\caption{Principal component analysis of VQE solution points~$\boldsymbol\theta$ for $m=-0.5$ (a, b), $m=0$ (c, d), and $m=1$ (e, f). Left figures correspond to ``top'' view and right ones to ``side'' projection. Color shows target function values $E(\boldsymbol\theta)$.}
\label{fig:PCA}
\end{figure*}

Fig.~\ref{fig:PCA} presents obtained PCA projections for $m = -0.5$, $0$, and $1$ in two views, which will be called ``top'' and ``side'' for convenience. Color shows target function values $E(\boldsymbol\theta)$. Blue points correspond to erroneous eigenvalues for the given $m$. The overall structure is similar for different $m$, especially for $-0.5$ and $0$. However, the majority of converged values are not correct for $m = 0$. For $m = 1$, the fraction of good solutions increases, blue areas slowly disappear. This suggests that the algorithm converges to the desired point with higher probability, which is in perfect agreement with the histogram in Fig.~\ref{fig:hist_m1}. There are two classes of proper solutions: points from the first class are located ``inside'' the erroneous plateau and for the second lie ``outside''. However, it appears difficult to isolate areas of initial points~$\boldsymbol\theta_0$ that can guarantee finding the true minimum or lead to one or another class of solutions, and additional research is required.

\clearpage

%% file: aipsamp.bbl
\providecommand{\noopsort}[1]{}\providecommand{\singleletter}[1]{#1}%
\begin{thebibliography}{36}%
\makeatletter
\providecommand \@ifxundefined [1]{%
 \@ifx{#1\undefined}
}%
\providecommand \@ifnum [1]{%
 \ifnum #1\expandafter \@firstoftwo
 \else \expandafter \@secondoftwo
 \fi
}%
\providecommand \@ifx [1]{%
 \ifx #1\expandafter \@firstoftwo
 \else \expandafter \@secondoftwo
 \fi
}%
\providecommand \natexlab [1]{#1}%
\providecommand \enquote  [1]{``#1''}%
\providecommand \bibnamefont  [1]{#1}%
\providecommand \bibfnamefont [1]{#1}%
\providecommand \citenamefont [1]{#1}%
\providecommand \href@noop [0]{\@secondoftwo}%
\providecommand \href [0]{\begingroup \@sanitize@url \@href}%
\providecommand \@href[1]{\@@startlink{#1}\@@href}%
\providecommand \@@href[1]{\endgroup#1\@@endlink}%
\providecommand \@sanitize@url [0]{\catcode `\\12\catcode `\$12\catcode
  `\&12\catcode `\#12\catcode `\^12\catcode `\_12\catcode `\%12\relax}%
\providecommand \@@startlink[1]{}%
\providecommand \@@endlink[0]{}%
\providecommand \url  [0]{\begingroup\@sanitize@url \@url }%
\providecommand \@url [1]{\endgroup\@href {#1}{\urlprefix }}%
\providecommand \urlprefix  [0]{URL }%
\providecommand \Eprint [0]{\href }%
\providecommand \doibase [0]{http://dx.doi.org/}%
\providecommand \selectlanguage [0]{\@gobble}%
\providecommand \bibinfo  [0]{\@secondoftwo}%
\providecommand \bibfield  [0]{\@secondoftwo}%
\providecommand \translation [1]{[#1]}%
\providecommand \BibitemOpen [0]{}%
\providecommand \bibitemStop [0]{}%
\providecommand \bibitemNoStop [0]{.\EOS\space}%
\providecommand \EOS [0]{\spacefactor3000\relax}%
\providecommand \BibitemShut  [1]{\csname bibitem#1\endcsname}%
\let\auto@bib@innerbib\@empty
\bibitem [{\citenamefont {Lambert}\ \emph {et~al.}(2013)\citenamefont
  {Lambert}, \citenamefont {Chen}, \citenamefont {Cheng}, \citenamefont {Li},
  \citenamefont {Chen},\ and\ \citenamefont {Nori}}]{lambert2013quantum}%
  \BibitemOpen
  \bibfield  {author} {\bibinfo {author} {\bibfnamefont {N.}~\bibnamefont
  {Lambert}}, \bibinfo {author} {\bibfnamefont {Y.-N.}\ \bibnamefont {Chen}},
  \bibinfo {author} {\bibfnamefont {Y.-C.}\ \bibnamefont {Cheng}}, \bibinfo
  {author} {\bibfnamefont {C.-M.}\ \bibnamefont {Li}}, \bibinfo {author}
  {\bibfnamefont {G.-Y.}\ \bibnamefont {Chen}}, \ and\ \bibinfo {author}
  {\bibfnamefont {F.}~\bibnamefont {Nori}},\ }\href@noop {} {\bibfield
  {journal} {\bibinfo  {journal} {Nature Physics}\ }\textbf {\bibinfo {volume}
  {9}},\ \bibinfo {pages} {10} (\bibinfo {year} {2013})}\BibitemShut {NoStop}%
\bibitem [{\citenamefont {Neukart}\ \emph {et~al.}(2017)\citenamefont
  {Neukart}, \citenamefont {Compostella}, \citenamefont {Seidel}, \citenamefont
  {Von~Dollen}, \citenamefont {Yarkoni},\ and\ \citenamefont
  {Parney}}]{neukart2017traffic}%
  \BibitemOpen
  \bibfield  {author} {\bibinfo {author} {\bibfnamefont {F.}~\bibnamefont
  {Neukart}}, \bibinfo {author} {\bibfnamefont {G.}~\bibnamefont
  {Compostella}}, \bibinfo {author} {\bibfnamefont {C.}~\bibnamefont {Seidel}},
  \bibinfo {author} {\bibfnamefont {D.}~\bibnamefont {Von~Dollen}}, \bibinfo
  {author} {\bibfnamefont {S.}~\bibnamefont {Yarkoni}}, \ and\ \bibinfo
  {author} {\bibfnamefont {B.}~\bibnamefont {Parney}},\ }\href@noop {}
  {\bibfield  {journal} {\bibinfo  {journal} {Frontiers in ICT}\ }\textbf
  {\bibinfo {volume} {4}},\ \bibinfo {pages} {29} (\bibinfo {year}
  {2017})}\BibitemShut {NoStop}%
\bibitem [{\citenamefont {Werlang}\ \emph {et~al.}(2010)\citenamefont
  {Werlang}, \citenamefont {Trippe}, \citenamefont {Ribeiro},\ and\
  \citenamefont {Rigolin}}]{werlang2010quantum}%
  \BibitemOpen
  \bibfield  {author} {\bibinfo {author} {\bibfnamefont {T.}~\bibnamefont
  {Werlang}}, \bibinfo {author} {\bibfnamefont {C.}~\bibnamefont {Trippe}},
  \bibinfo {author} {\bibfnamefont {G.}~\bibnamefont {Ribeiro}}, \ and\
  \bibinfo {author} {\bibfnamefont {G.}~\bibnamefont {Rigolin}},\ }\href@noop
  {} {\bibfield  {journal} {\bibinfo  {journal} {Physical review letters}\
  }\textbf {\bibinfo {volume} {105}},\ \bibinfo {pages} {095702} (\bibinfo
  {year} {2010})}\BibitemShut {NoStop}%
\bibitem [{\citenamefont {Abadie}\ \emph {et~al.}(2011)\citenamefont {Abadie},
  \citenamefont {Abbott}, \citenamefont {Abbott}, \citenamefont {Abbott},
  \citenamefont {Abernathy}, \citenamefont {Adams}, \citenamefont {Adhikari},
  \citenamefont {Affeldt}, \citenamefont {Allen}, \citenamefont {Allen} \emph
  {et~al.}}]{abadie2011gravitational}%
  \BibitemOpen
  \bibfield  {author} {\bibinfo {author} {\bibfnamefont {J.}~\bibnamefont
  {Abadie}}, \bibinfo {author} {\bibfnamefont {B.~P.}\ \bibnamefont {Abbott}},
  \bibinfo {author} {\bibfnamefont {R.}~\bibnamefont {Abbott}}, \bibinfo
  {author} {\bibfnamefont {T.~D.}\ \bibnamefont {Abbott}}, \bibinfo {author}
  {\bibfnamefont {M.}~\bibnamefont {Abernathy}}, \bibinfo {author}
  {\bibfnamefont {C.}~\bibnamefont {Adams}}, \bibinfo {author} {\bibfnamefont
  {R.}~\bibnamefont {Adhikari}}, \bibinfo {author} {\bibfnamefont
  {C.}~\bibnamefont {Affeldt}}, \bibinfo {author} {\bibfnamefont
  {B.}~\bibnamefont {Allen}}, \bibinfo {author} {\bibfnamefont
  {G.}~\bibnamefont {Allen}},  \emph {et~al.},\ }\href@noop {} {\bibfield
  {journal} {\bibinfo  {journal} {Nature Physics}\ }\textbf {\bibinfo {volume}
  {7}},\ \bibinfo {pages} {962} (\bibinfo {year} {2011})}\BibitemShut {NoStop}%
\bibitem [{\citenamefont {Lubasch}\ \emph {et~al.}(2020)\citenamefont
  {Lubasch}, \citenamefont {Joo}, \citenamefont {Moinier}, \citenamefont
  {Kiffner},\ and\ \citenamefont {Jaksch}}]{lubasch2020variational}%
  \BibitemOpen
  \bibfield  {author} {\bibinfo {author} {\bibfnamefont {M.}~\bibnamefont
  {Lubasch}}, \bibinfo {author} {\bibfnamefont {J.}~\bibnamefont {Joo}},
  \bibinfo {author} {\bibfnamefont {P.}~\bibnamefont {Moinier}}, \bibinfo
  {author} {\bibfnamefont {M.}~\bibnamefont {Kiffner}}, \ and\ \bibinfo
  {author} {\bibfnamefont {D.}~\bibnamefont {Jaksch}},\ }\href@noop {}
  {\bibfield  {journal} {\bibinfo  {journal} {Physical Review A}\ }\textbf
  {\bibinfo {volume} {101}},\ \bibinfo {pages} {010301} (\bibinfo {year}
  {2020})}\BibitemShut {NoStop}%
\bibitem [{\citenamefont {Feynman}(1986)}]{Feynman1986}%
  \BibitemOpen
  \bibfield  {author} {\bibinfo {author} {\bibfnamefont {R.~P.}\ \bibnamefont
  {Feynman}},\ }\href {\doibase 10.1007/BF01886518} {\bibfield  {journal}
  {\bibinfo  {journal} {Foundations of Physics}\ }\textbf {\bibinfo {volume}
  {16}},\ \bibinfo {pages} {507} (\bibinfo {year} {1986})}\BibitemShut
  {NoStop}%
\bibitem [{\citenamefont {Peruzzo}\ \emph {et~al.}(2014)\citenamefont
  {Peruzzo}, \citenamefont {McClean}, \citenamefont {Shadbolt}, \citenamefont
  {Yung}, \citenamefont {Zhou}, \citenamefont {Love}, \citenamefont
  {Aspuru-Guzik},\ and\ \citenamefont {O’brien}}]{obrien2014vqe}%
  \BibitemOpen
  \bibfield  {author} {\bibinfo {author} {\bibfnamefont {A.}~\bibnamefont
  {Peruzzo}}, \bibinfo {author} {\bibfnamefont {J.}~\bibnamefont {McClean}},
  \bibinfo {author} {\bibfnamefont {P.}~\bibnamefont {Shadbolt}}, \bibinfo
  {author} {\bibfnamefont {M.-H.}\ \bibnamefont {Yung}}, \bibinfo {author}
  {\bibfnamefont {X.-Q.}\ \bibnamefont {Zhou}}, \bibinfo {author}
  {\bibfnamefont {P.~J.}\ \bibnamefont {Love}}, \bibinfo {author}
  {\bibfnamefont {A.}~\bibnamefont {Aspuru-Guzik}}, \ and\ \bibinfo {author}
  {\bibfnamefont {J.~L.}\ \bibnamefont {O’brien}},\ }\href@noop {} {\bibfield
   {journal} {\bibinfo  {journal} {Nature communications}\ }\textbf {\bibinfo
  {volume} {5}},\ \bibinfo {pages} {4213} (\bibinfo {year} {2014})}\BibitemShut
  {NoStop}%
\bibitem [{\citenamefont {Yung}\ \emph {et~al.}(2014)\citenamefont {Yung},
  \citenamefont {Casanova}, \citenamefont {Mezzacapo}, \citenamefont {Mcclean},
  \citenamefont {Lamata}, \citenamefont {Aspuru-Guzik},\ and\ \citenamefont
  {Solano}}]{yung2014transistor}%
  \BibitemOpen
  \bibfield  {author} {\bibinfo {author} {\bibfnamefont {M.-H.}\ \bibnamefont
  {Yung}}, \bibinfo {author} {\bibfnamefont {J.}~\bibnamefont {Casanova}},
  \bibinfo {author} {\bibfnamefont {A.}~\bibnamefont {Mezzacapo}}, \bibinfo
  {author} {\bibfnamefont {J.}~\bibnamefont {Mcclean}}, \bibinfo {author}
  {\bibfnamefont {L.}~\bibnamefont {Lamata}}, \bibinfo {author} {\bibfnamefont
  {A.}~\bibnamefont {Aspuru-Guzik}}, \ and\ \bibinfo {author} {\bibfnamefont
  {E.}~\bibnamefont {Solano}},\ }\href@noop {} {\bibfield  {journal} {\bibinfo
  {journal} {Scientific Reports}\ }\textbf {\bibinfo {volume} {4}},\ \bibinfo
  {pages} {3589} (\bibinfo {year} {2014})}\BibitemShut {NoStop}%
\bibitem [{\citenamefont {Shen}\ \emph {et~al.}(2017)\citenamefont {Shen},
  \citenamefont {Zhang}, \citenamefont {Zhang}, \citenamefont {Zhang},
  \citenamefont {Yung},\ and\ \citenamefont {Kim}}]{shen2017quantum}%
  \BibitemOpen
  \bibfield  {author} {\bibinfo {author} {\bibfnamefont {Y.}~\bibnamefont
  {Shen}}, \bibinfo {author} {\bibfnamefont {X.}~\bibnamefont {Zhang}},
  \bibinfo {author} {\bibfnamefont {S.}~\bibnamefont {Zhang}}, \bibinfo
  {author} {\bibfnamefont {J.-N.}\ \bibnamefont {Zhang}}, \bibinfo {author}
  {\bibfnamefont {M.-H.}\ \bibnamefont {Yung}}, \ and\ \bibinfo {author}
  {\bibfnamefont {K.}~\bibnamefont {Kim}},\ }\href@noop {} {\bibfield
  {journal} {\bibinfo  {journal} {Physical Review A}\ }\textbf {\bibinfo
  {volume} {95}},\ \bibinfo {pages} {020501} (\bibinfo {year}
  {2017})}\BibitemShut {NoStop}%
\bibitem [{\citenamefont {O’Malley}\ \emph {et~al.}(2016)\citenamefont
  {O’Malley}, \citenamefont {Babbush}, \citenamefont {Kivlichan},
  \citenamefont {Romero}, \citenamefont {McClean}, \citenamefont {Barends},
  \citenamefont {Kelly}, \citenamefont {Roushan}, \citenamefont {Tranter},
  \citenamefont {Ding} \emph {et~al.}}]{o2016scalable}%
  \BibitemOpen
  \bibfield  {author} {\bibinfo {author} {\bibfnamefont {P.~J.}\ \bibnamefont
  {O’Malley}}, \bibinfo {author} {\bibfnamefont {R.}~\bibnamefont {Babbush}},
  \bibinfo {author} {\bibfnamefont {I.~D.}\ \bibnamefont {Kivlichan}}, \bibinfo
  {author} {\bibfnamefont {J.}~\bibnamefont {Romero}}, \bibinfo {author}
  {\bibfnamefont {J.~R.}\ \bibnamefont {McClean}}, \bibinfo {author}
  {\bibfnamefont {R.}~\bibnamefont {Barends}}, \bibinfo {author} {\bibfnamefont
  {J.}~\bibnamefont {Kelly}}, \bibinfo {author} {\bibfnamefont
  {P.}~\bibnamefont {Roushan}}, \bibinfo {author} {\bibfnamefont
  {A.}~\bibnamefont {Tranter}}, \bibinfo {author} {\bibfnamefont
  {N.}~\bibnamefont {Ding}},  \emph {et~al.},\ }\href@noop {} {\bibfield
  {journal} {\bibinfo  {journal} {Physical Review X}\ }\textbf {\bibinfo
  {volume} {6}},\ \bibinfo {pages} {031007} (\bibinfo {year}
  {2016})}\BibitemShut {NoStop}%
\bibitem [{\citenamefont {Kandala}\ \emph {et~al.}(2017)\citenamefont
  {Kandala}, \citenamefont {Mezzacapo}, \citenamefont {Temme}, \citenamefont
  {Takita}, \citenamefont {Brink}, \citenamefont {Chow},\ and\ \citenamefont
  {Gambetta}}]{kandala2017hardware}%
  \BibitemOpen
  \bibfield  {author} {\bibinfo {author} {\bibfnamefont {A.}~\bibnamefont
  {Kandala}}, \bibinfo {author} {\bibfnamefont {A.}~\bibnamefont {Mezzacapo}},
  \bibinfo {author} {\bibfnamefont {K.}~\bibnamefont {Temme}}, \bibinfo
  {author} {\bibfnamefont {M.}~\bibnamefont {Takita}}, \bibinfo {author}
  {\bibfnamefont {M.}~\bibnamefont {Brink}}, \bibinfo {author} {\bibfnamefont
  {J.~M.}\ \bibnamefont {Chow}}, \ and\ \bibinfo {author} {\bibfnamefont
  {J.~M.}\ \bibnamefont {Gambetta}},\ }\href@noop {} {\bibfield  {journal}
  {\bibinfo  {journal} {Nature}\ }\textbf {\bibinfo {volume} {549}},\ \bibinfo
  {pages} {242} (\bibinfo {year} {2017})}\BibitemShut {NoStop}%
\bibitem [{\citenamefont {Kokail}\ \emph {et~al.}(2019)\citenamefont {Kokail},
  \citenamefont {Maier}, \citenamefont {van Bijnen}, \citenamefont {Brydges},
  \citenamefont {Joshi}, \citenamefont {Jurcevic}, \citenamefont {Muschik},
  \citenamefont {Silvi}, \citenamefont {Blatt}, \citenamefont {Roos} \emph
  {et~al.}}]{kokail2019self}%
  \BibitemOpen
  \bibfield  {author} {\bibinfo {author} {\bibfnamefont {C.}~\bibnamefont
  {Kokail}}, \bibinfo {author} {\bibfnamefont {C.}~\bibnamefont {Maier}},
  \bibinfo {author} {\bibfnamefont {R.}~\bibnamefont {van Bijnen}}, \bibinfo
  {author} {\bibfnamefont {T.}~\bibnamefont {Brydges}}, \bibinfo {author}
  {\bibfnamefont {M.~K.}\ \bibnamefont {Joshi}}, \bibinfo {author}
  {\bibfnamefont {P.}~\bibnamefont {Jurcevic}}, \bibinfo {author}
  {\bibfnamefont {C.~A.}\ \bibnamefont {Muschik}}, \bibinfo {author}
  {\bibfnamefont {P.}~\bibnamefont {Silvi}}, \bibinfo {author} {\bibfnamefont
  {R.}~\bibnamefont {Blatt}}, \bibinfo {author} {\bibfnamefont {C.~F.}\
  \bibnamefont {Roos}},  \emph {et~al.},\ }\href@noop {} {\bibfield  {journal}
  {\bibinfo  {journal} {Nature}\ }\textbf {\bibinfo {volume} {569}},\ \bibinfo
  {pages} {355} (\bibinfo {year} {2019})}\BibitemShut {NoStop}%
\bibitem [{\citenamefont {Uvarov}\ \emph {et~al.}(2020)\citenamefont {Uvarov},
  \citenamefont {Kardashin},\ and\ \citenamefont {Biamonte}}]{Uvarov2020}%
  \BibitemOpen
  \bibfield  {author} {\bibinfo {author} {\bibfnamefont {A.~V.}\ \bibnamefont
  {Uvarov}}, \bibinfo {author} {\bibfnamefont {A.~S.}\ \bibnamefont
  {Kardashin}}, \ and\ \bibinfo {author} {\bibfnamefont {J.~D.}\ \bibnamefont
  {Biamonte}},\ }\href {\doibase 10.1103/physreva.102.012415} {\bibfield
  {journal} {\bibinfo  {journal} {Physical Review A}\ }\textbf {\bibinfo
  {volume} {102}} (\bibinfo {year} {2020}),\
  10.1103/physreva.102.012415}\BibitemShut {NoStop}%
\bibitem [{\citenamefont {Wang}\ \emph {et~al.}(2019)\citenamefont {Wang},
  \citenamefont {Higgott},\ and\ \citenamefont
  {Brierley}}]{wang2019accelerated}%
  \BibitemOpen
  \bibfield  {author} {\bibinfo {author} {\bibfnamefont {D.}~\bibnamefont
  {Wang}}, \bibinfo {author} {\bibfnamefont {O.}~\bibnamefont {Higgott}}, \
  and\ \bibinfo {author} {\bibfnamefont {S.}~\bibnamefont {Brierley}},\
  }\href@noop {} {\bibfield  {journal} {\bibinfo  {journal} {Physical review
  letters}\ }\textbf {\bibinfo {volume} {122}},\ \bibinfo {pages} {140504}
  (\bibinfo {year} {2019})}\BibitemShut {NoStop}%
\bibitem [{\citenamefont {{Biamonte}}\ \emph {et~al.}(2017)\citenamefont
  {{Biamonte}}, \citenamefont {{Wittek}}, \citenamefont {{Pancotti}},
  \citenamefont {{Rebentrost}}, \citenamefont {{Wiebe}},\ and\ \citenamefont
  {{Lloyd}}}]{2017Natur.549..195B}%
  \BibitemOpen
  \bibfield  {author} {\bibinfo {author} {\bibfnamefont {J.}~\bibnamefont
  {{Biamonte}}}, \bibinfo {author} {\bibfnamefont {P.}~\bibnamefont
  {{Wittek}}}, \bibinfo {author} {\bibfnamefont {N.}~\bibnamefont
  {{Pancotti}}}, \bibinfo {author} {\bibfnamefont {P.}~\bibnamefont
  {{Rebentrost}}}, \bibinfo {author} {\bibfnamefont {N.}~\bibnamefont
  {{Wiebe}}}, \ and\ \bibinfo {author} {\bibfnamefont {S.}~\bibnamefont
  {{Lloyd}}},\ }\href {\doibase 10.1038/nature23474} {\bibfield  {journal}
  {\bibinfo  {journal} {Nature}\ }\textbf {\bibinfo {volume} {549}},\ \bibinfo
  {pages} {195} (\bibinfo {year} {2017})},\ \Eprint
  {http://arxiv.org/abs/1611.09347} {arXiv:1611.09347 [quant-ph]} \BibitemShut
  {NoStop}%
\bibitem [{\citenamefont {Akshay}\ \emph {et~al.}(2020)\citenamefont {Akshay},
  \citenamefont {Philathong}, \citenamefont {Morales},\ and\ \citenamefont
  {Biamonte}}]{Akshay_2020}%
  \BibitemOpen
  \bibfield  {author} {\bibinfo {author} {\bibfnamefont {V.}~\bibnamefont
  {Akshay}}, \bibinfo {author} {\bibfnamefont {H.}~\bibnamefont {Philathong}},
  \bibinfo {author} {\bibfnamefont {M.}~\bibnamefont {Morales}}, \ and\
  \bibinfo {author} {\bibfnamefont {J.}~\bibnamefont {Biamonte}},\ }\href
  {\doibase 10.1103/physrevlett.124.090504} {\bibfield  {journal} {\bibinfo
  {journal} {Physical Review Letters}\ }\textbf {\bibinfo {volume} {124}}
  (\bibinfo {year} {2020}),\ 10.1103/physrevlett.124.090504}\BibitemShut
  {NoStop}%
\bibitem [{\citenamefont {Farhi}\ \emph {et~al.}(2014)\citenamefont {Farhi},
  \citenamefont {Goldstone},\ and\ \citenamefont {Gutmann}}]{farhi2014quantum}%
  \BibitemOpen
  \bibfield  {author} {\bibinfo {author} {\bibfnamefont {E.}~\bibnamefont
  {Farhi}}, \bibinfo {author} {\bibfnamefont {J.}~\bibnamefont {Goldstone}}, \
  and\ \bibinfo {author} {\bibfnamefont {S.}~\bibnamefont {Gutmann}},\
  }\href@noop {} {\enquote {\bibinfo {title} {A quantum approximate
  optimization algorithm},}\ } (\bibinfo {year} {2014}),\ \bibinfo {note}
  {unpublished, arXiv:1411.4028}\BibitemShut {NoStop}%
\bibitem [{\citenamefont {Mitarai}\ \emph {et~al.}(2019)\citenamefont
  {Mitarai}, \citenamefont {Yan},\ and\ \citenamefont
  {Fujii}}]{mitarai2019generalization}%
  \BibitemOpen
  \bibfield  {author} {\bibinfo {author} {\bibfnamefont {K.}~\bibnamefont
  {Mitarai}}, \bibinfo {author} {\bibfnamefont {T.}~\bibnamefont {Yan}}, \ and\
  \bibinfo {author} {\bibfnamefont {K.}~\bibnamefont {Fujii}},\ }\href@noop {}
  {\bibfield  {journal} {\bibinfo  {journal} {Physical Review Applied}\
  }\textbf {\bibinfo {volume} {11}},\ \bibinfo {pages} {044087} (\bibinfo
  {year} {2019})}\BibitemShut {NoStop}%
\bibitem [{\citenamefont {Biamonte}(2019)}]{biamonte2019universal}%
  \BibitemOpen
  \bibfield  {author} {\bibinfo {author} {\bibfnamefont {J.}~\bibnamefont
  {Biamonte}},\ }\href@noop {} {\enquote {\bibinfo {title} {Universal
  variational quantum computation},}\ } (\bibinfo {year} {2019}),\ \Eprint
  {http://arxiv.org/abs/1903.04500} {arXiv:1903.04500 [quant-ph]} \BibitemShut
  {NoStop}%
\bibitem [{\citenamefont {Morales}\ \emph {et~al.}(2020)\citenamefont
  {Morales}, \citenamefont {Biamonte},\ and\ \citenamefont
  {Zimbor{\'{a}}s}}]{morales2019universality}%
  \BibitemOpen
  \bibfield  {author} {\bibinfo {author} {\bibfnamefont {M.~E.~S.}\
  \bibnamefont {Morales}}, \bibinfo {author} {\bibfnamefont {J.~D.}\
  \bibnamefont {Biamonte}}, \ and\ \bibinfo {author} {\bibfnamefont
  {Z.}~\bibnamefont {Zimbor{\'{a}}s}},\ }\href {\doibase
  10.1007/s11128-020-02748-9} {\bibfield  {journal} {\bibinfo  {journal}
  {Quantum Information Processing}\ }\textbf {\bibinfo {volume} {19}} (\bibinfo
  {year} {2020}),\ 10.1007/s11128-020-02748-9}\BibitemShut {NoStop}%
\bibitem [{\citenamefont {Carolan}\ \emph {et~al.}(2020)\citenamefont
  {Carolan}, \citenamefont {Mohseni}, \citenamefont {Olson}, \citenamefont
  {Prabhu}, \citenamefont {Chen}, \citenamefont {Bunandar}, \citenamefont
  {Niu}, \citenamefont {Harris}, \citenamefont {Wong}, \citenamefont {Hochberg}
  \emph {et~al.}}]{carolan2020variational}%
  \BibitemOpen
  \bibfield  {author} {\bibinfo {author} {\bibfnamefont {J.}~\bibnamefont
  {Carolan}}, \bibinfo {author} {\bibfnamefont {M.}~\bibnamefont {Mohseni}},
  \bibinfo {author} {\bibfnamefont {J.~P.}\ \bibnamefont {Olson}}, \bibinfo
  {author} {\bibfnamefont {M.}~\bibnamefont {Prabhu}}, \bibinfo {author}
  {\bibfnamefont {C.}~\bibnamefont {Chen}}, \bibinfo {author} {\bibfnamefont
  {D.}~\bibnamefont {Bunandar}}, \bibinfo {author} {\bibfnamefont {M.~Y.}\
  \bibnamefont {Niu}}, \bibinfo {author} {\bibfnamefont {N.~C.}\ \bibnamefont
  {Harris}}, \bibinfo {author} {\bibfnamefont {F.~N.}\ \bibnamefont {Wong}},
  \bibinfo {author} {\bibfnamefont {M.}~\bibnamefont {Hochberg}},  \emph
  {et~al.},\ }\href@noop {} {\bibfield  {journal} {\bibinfo  {journal} {Nature
  Physics}\ }\textbf {\bibinfo {volume} {16}},\ \bibinfo {pages} {322}
  (\bibinfo {year} {2020})}\BibitemShut {NoStop}%
\bibitem [{\citenamefont {Ryabinkin}\ \emph {et~al.}(2018)\citenamefont
  {Ryabinkin}, \citenamefont {Genin},\ and\ \citenamefont
  {Izmaylov}}]{ryabinkin2018constrained}%
  \BibitemOpen
  \bibfield  {author} {\bibinfo {author} {\bibfnamefont {I.~G.}\ \bibnamefont
  {Ryabinkin}}, \bibinfo {author} {\bibfnamefont {S.~N.}\ \bibnamefont
  {Genin}}, \ and\ \bibinfo {author} {\bibfnamefont {A.~F.}\ \bibnamefont
  {Izmaylov}},\ }\href@noop {} {\bibfield  {journal} {\bibinfo  {journal}
  {Journal of chemical theory and computation}\ }\textbf {\bibinfo {volume}
  {15}},\ \bibinfo {pages} {249} (\bibinfo {year} {2018})}\BibitemShut
  {NoStop}%
\bibitem [{\citenamefont {Parrish}\ \emph {et~al.}(2019)\citenamefont
  {Parrish}, \citenamefont {Hohenstein}, \citenamefont {McMahon},\ and\
  \citenamefont {Mart{\'\i}nez}}]{parrish2019quantum}%
  \BibitemOpen
  \bibfield  {author} {\bibinfo {author} {\bibfnamefont {R.~M.}\ \bibnamefont
  {Parrish}}, \bibinfo {author} {\bibfnamefont {E.~G.}\ \bibnamefont
  {Hohenstein}}, \bibinfo {author} {\bibfnamefont {P.~L.}\ \bibnamefont
  {McMahon}}, \ and\ \bibinfo {author} {\bibfnamefont {T.~J.}\ \bibnamefont
  {Mart{\'\i}nez}},\ }\href@noop {} {\bibfield  {journal} {\bibinfo  {journal}
  {Physical review letters}\ }\textbf {\bibinfo {volume} {122}},\ \bibinfo
  {pages} {230401} (\bibinfo {year} {2019})}\BibitemShut {NoStop}%
\bibitem [{\citenamefont {McClean}\ \emph {et~al.}(2018)\citenamefont
  {McClean}, \citenamefont {Boixo}, \citenamefont {Smelyanskiy}, \citenamefont
  {Babbush},\ and\ \citenamefont {Neven}}]{McClean_2018}%
  \BibitemOpen
  \bibfield  {author} {\bibinfo {author} {\bibfnamefont {J.~R.}\ \bibnamefont
  {McClean}}, \bibinfo {author} {\bibfnamefont {S.}~\bibnamefont {Boixo}},
  \bibinfo {author} {\bibfnamefont {V.~N.}\ \bibnamefont {Smelyanskiy}},
  \bibinfo {author} {\bibfnamefont {R.}~\bibnamefont {Babbush}}, \ and\
  \bibinfo {author} {\bibfnamefont {H.}~\bibnamefont {Neven}},\ }\href
  {\doibase 10.1038/s41467-018-07090-4} {\bibfield  {journal} {\bibinfo
  {journal} {Nature Communications}\ }\textbf {\bibinfo {volume} {9}} (\bibinfo
  {year} {2018}),\ 10.1038/s41467-018-07090-4}\BibitemShut {NoStop}%
\bibitem [{\citenamefont {Higgott}\ \emph {et~al.}(2019)\citenamefont
  {Higgott}, \citenamefont {Wang},\ and\ \citenamefont
  {Brierley}}]{higgott2019variational}%
  \BibitemOpen
  \bibfield  {author} {\bibinfo {author} {\bibfnamefont {O.}~\bibnamefont
  {Higgott}}, \bibinfo {author} {\bibfnamefont {D.}~\bibnamefont {Wang}}, \
  and\ \bibinfo {author} {\bibfnamefont {S.}~\bibnamefont {Brierley}},\
  }\href@noop {} {\bibfield  {journal} {\bibinfo  {journal} {Quantum}\ }\textbf
  {\bibinfo {volume} {3}},\ \bibinfo {pages} {156} (\bibinfo {year}
  {2019})}\BibitemShut {NoStop}%
\bibitem [{\citenamefont {Pechen}(2011)}]{Pechen2011}%
  \BibitemOpen
  \bibfield  {author} {\bibinfo {author} {\bibfnamefont {A.}~\bibnamefont
  {Pechen}},\ }\href {\doibase 10.1103/physreva.84.042106} {\bibfield
  {journal} {\bibinfo  {journal} {Physical Review A}\ }\textbf {\bibinfo
  {volume} {84}} (\bibinfo {year} {2011}),\
  10.1103/physreva.84.042106}\BibitemShut {NoStop}%
\bibitem [{\citenamefont {Fedrizzi}\ \emph {et~al.}(2007)\citenamefont
  {Fedrizzi}, \citenamefont {Herbst}, \citenamefont {Poppe}, \citenamefont
  {Jennewein},\ and\ \citenamefont {Zeilinger}}]{zeilinger2007source}%
  \BibitemOpen
  \bibfield  {author} {\bibinfo {author} {\bibfnamefont {A.}~\bibnamefont
  {Fedrizzi}}, \bibinfo {author} {\bibfnamefont {T.}~\bibnamefont {Herbst}},
  \bibinfo {author} {\bibfnamefont {A.}~\bibnamefont {Poppe}}, \bibinfo
  {author} {\bibfnamefont {T.}~\bibnamefont {Jennewein}}, \ and\ \bibinfo
  {author} {\bibfnamefont {A.}~\bibnamefont {Zeilinger}},\ }\href {\doibase
  10.1364/OE.15.015377} {\bibfield  {journal} {\bibinfo  {journal} {Optics
  Express}\ }\textbf {\bibinfo {volume} {15}},\ \bibinfo {pages} {15377}
  (\bibinfo {year} {2007})}\BibitemShut {NoStop}%
\bibitem [{\citenamefont {Byrnes}\ \emph {et~al.}(2002)\citenamefont {Byrnes},
  \citenamefont {Sriganesh}, \citenamefont {Bursill},\ and\ \citenamefont
  {Hamer}}]{PhysRevD.66.013002}%
  \BibitemOpen
  \bibfield  {author} {\bibinfo {author} {\bibfnamefont {T.~M.~R.}\
  \bibnamefont {Byrnes}}, \bibinfo {author} {\bibfnamefont {P.}~\bibnamefont
  {Sriganesh}}, \bibinfo {author} {\bibfnamefont {R.~J.}\ \bibnamefont
  {Bursill}}, \ and\ \bibinfo {author} {\bibfnamefont {C.~J.}\ \bibnamefont
  {Hamer}},\ }\href {\doibase 10.1103/PhysRevD.66.013002} {\bibfield  {journal}
  {\bibinfo  {journal} {Phys. Rev. D}\ }\textbf {\bibinfo {volume} {66}},\
  \bibinfo {pages} {013002} (\bibinfo {year} {2002})}\BibitemShut {NoStop}%
\bibitem [{\citenamefont {Byrnes}\ and\ \citenamefont
  {Yamamoto}(2006)}]{PhysRevA.73.022328}%
  \BibitemOpen
  \bibfield  {author} {\bibinfo {author} {\bibfnamefont {T.}~\bibnamefont
  {Byrnes}}\ and\ \bibinfo {author} {\bibfnamefont {Y.}~\bibnamefont
  {Yamamoto}},\ }\href {\doibase 10.1103/PhysRevA.73.022328} {\bibfield
  {journal} {\bibinfo  {journal} {Phys. Rev. A}\ }\textbf {\bibinfo {volume}
  {73}},\ \bibinfo {pages} {022328} (\bibinfo {year} {2006})}\BibitemShut
  {NoStop}%
\bibitem [{\citenamefont {Rosenbrock}(1960)}]{Rosenbrock_CJ1960}%
  \BibitemOpen
  \bibfield  {author} {\bibinfo {author} {\bibfnamefont {H.~H.}\ \bibnamefont
  {Rosenbrock}},\ }\href {\doibase 10.1093/comjnl/3.3.175} {\bibfield
  {journal} {\bibinfo  {journal} {The Computer Journal}\ }\textbf {\bibinfo
  {volume} {3}},\ \bibinfo {pages} {175} (\bibinfo {year} {1960})}\BibitemShut
  {NoStop}%
\bibitem [{\citenamefont {Bravo-Prieto}\ \emph {et~al.}(2020)\citenamefont
  {Bravo-Prieto}, \citenamefont {Lumbreras-Zarapico}, \citenamefont
  {Tagliacozzo},\ and\ \citenamefont {Latorre}}]{bravo2020scaling}%
  \BibitemOpen
  \bibfield  {author} {\bibinfo {author} {\bibfnamefont {C.}~\bibnamefont
  {Bravo-Prieto}}, \bibinfo {author} {\bibfnamefont {J.}~\bibnamefont
  {Lumbreras-Zarapico}}, \bibinfo {author} {\bibfnamefont {L.}~\bibnamefont
  {Tagliacozzo}}, \ and\ \bibinfo {author} {\bibfnamefont {J.~I.}\ \bibnamefont
  {Latorre}},\ }\href@noop {} {\bibfield  {journal} {\bibinfo  {journal}
  {Quantum}\ }\textbf {\bibinfo {volume} {4}},\ \bibinfo {pages} {272}
  (\bibinfo {year} {2020})}\BibitemShut {NoStop}%
\bibitem [{\citenamefont {Wang}\ \emph {et~al.}(2016)\citenamefont {Wang},
  \citenamefont {Chen}, \citenamefont {Li}, \citenamefont {Huang},
  \citenamefont {Liu}, \citenamefont {Chen}, \citenamefont {Luo}, \citenamefont
  {Su}, \citenamefont {Wu}, \citenamefont {Li}, \citenamefont {Lu},
  \citenamefont {Hu}, \citenamefont {Jiang}, \citenamefont {Peng},
  \citenamefont {Li}, \citenamefont {Liu}, \citenamefont {Chen}, \citenamefont
  {Lu},\ and\ \citenamefont {Pan}}]{wang2016experimental}%
  \BibitemOpen
  \bibfield  {author} {\bibinfo {author} {\bibfnamefont {X.-L.}\ \bibnamefont
  {Wang}}, \bibinfo {author} {\bibfnamefont {L.-K.}\ \bibnamefont {Chen}},
  \bibinfo {author} {\bibfnamefont {W.}~\bibnamefont {Li}}, \bibinfo {author}
  {\bibfnamefont {H.-L.}\ \bibnamefont {Huang}}, \bibinfo {author}
  {\bibfnamefont {C.}~\bibnamefont {Liu}}, \bibinfo {author} {\bibfnamefont
  {C.}~\bibnamefont {Chen}}, \bibinfo {author} {\bibfnamefont {Y.-H.}\
  \bibnamefont {Luo}}, \bibinfo {author} {\bibfnamefont {Z.-E.}\ \bibnamefont
  {Su}}, \bibinfo {author} {\bibfnamefont {D.}~\bibnamefont {Wu}}, \bibinfo
  {author} {\bibfnamefont {Z.-D.}\ \bibnamefont {Li}}, \bibinfo {author}
  {\bibfnamefont {H.}~\bibnamefont {Lu}}, \bibinfo {author} {\bibfnamefont
  {Y.}~\bibnamefont {Hu}}, \bibinfo {author} {\bibfnamefont {X.}~\bibnamefont
  {Jiang}}, \bibinfo {author} {\bibfnamefont {C.-Z.}\ \bibnamefont {Peng}},
  \bibinfo {author} {\bibfnamefont {L.}~\bibnamefont {Li}}, \bibinfo {author}
  {\bibfnamefont {N.-L.}\ \bibnamefont {Liu}}, \bibinfo {author} {\bibfnamefont
  {Y.-A.}\ \bibnamefont {Chen}}, \bibinfo {author} {\bibfnamefont {C.-Y.}\
  \bibnamefont {Lu}}, \ and\ \bibinfo {author} {\bibfnamefont {J.-W.}\
  \bibnamefont {Pan}},\ }\href {\doibase 10.1103/PhysRevLett.117.210502}
  {\bibfield  {journal} {\bibinfo  {journal} {Phys. Rev. Lett.}\ }\textbf
  {\bibinfo {volume} {117}},\ \bibinfo {pages} {210502} (\bibinfo {year}
  {2016})}\BibitemShut {NoStop}%
\bibitem [{\citenamefont {Kok}\ \emph {et~al.}(2007)\citenamefont {Kok},
  \citenamefont {Munro}, \citenamefont {Nemoto}, \citenamefont {Ralph},
  \citenamefont {Dowling},\ and\ \citenamefont {Milburn}}]{Milburn_RevModPhys}%
  \BibitemOpen
  \bibfield  {author} {\bibinfo {author} {\bibfnamefont {P.}~\bibnamefont
  {Kok}}, \bibinfo {author} {\bibfnamefont {W.~J.}\ \bibnamefont {Munro}},
  \bibinfo {author} {\bibfnamefont {K.}~\bibnamefont {Nemoto}}, \bibinfo
  {author} {\bibfnamefont {T.~C.}\ \bibnamefont {Ralph}}, \bibinfo {author}
  {\bibfnamefont {J.~P.}\ \bibnamefont {Dowling}}, \ and\ \bibinfo {author}
  {\bibfnamefont {G.~J.}\ \bibnamefont {Milburn}},\ }\href {\doibase
  10.1103/RevModPhys.79.135} {\bibfield  {journal} {\bibinfo  {journal} {Rev.
  Mod. Phys.}\ }\textbf {\bibinfo {volume} {79}},\ \bibinfo {pages} {135}
  (\bibinfo {year} {2007})}\BibitemShut {NoStop}%
\bibitem [{\citenamefont {Bartolucci}\ \emph {et~al.}(2021)\citenamefont
  {Bartolucci}, \citenamefont {Birchall}, \citenamefont {Bombin}, \citenamefont
  {Cable}, \citenamefont {Dawson}, \citenamefont {Gimeno-Segovia},
  \citenamefont {Johnston}, \citenamefont {Kieling}, \citenamefont {Nickerson},
  \citenamefont {Pant} \emph {et~al.}}]{bartolucci2021fusion}%
  \BibitemOpen
  \bibfield  {author} {\bibinfo {author} {\bibfnamefont {S.}~\bibnamefont
  {Bartolucci}}, \bibinfo {author} {\bibfnamefont {P.}~\bibnamefont
  {Birchall}}, \bibinfo {author} {\bibfnamefont {H.}~\bibnamefont {Bombin}},
  \bibinfo {author} {\bibfnamefont {H.}~\bibnamefont {Cable}}, \bibinfo
  {author} {\bibfnamefont {C.}~\bibnamefont {Dawson}}, \bibinfo {author}
  {\bibfnamefont {M.}~\bibnamefont {Gimeno-Segovia}}, \bibinfo {author}
  {\bibfnamefont {E.}~\bibnamefont {Johnston}}, \bibinfo {author}
  {\bibfnamefont {K.}~\bibnamefont {Kieling}}, \bibinfo {author} {\bibfnamefont
  {N.}~\bibnamefont {Nickerson}}, \bibinfo {author} {\bibfnamefont
  {M.}~\bibnamefont {Pant}},  \emph {et~al.},\ }\href@noop {} {\bibfield
  {journal} {\bibinfo  {journal} {arXiv preprint arXiv:2101.09310}\ } (\bibinfo
  {year} {2021})}\BibitemShut {NoStop}%
\bibitem [{\citenamefont {{Spall}}(1992)}]{Spall_TAC1992}%
  \BibitemOpen
  \bibfield  {author} {\bibinfo {author} {\bibfnamefont {J.~C.}\ \bibnamefont
  {{Spall}}},\ }\href@noop {} {\bibfield  {journal} {\bibinfo  {journal} {IEEE
  Transactions on Automatic Control}\ }\textbf {\bibinfo {volume} {37}},\
  \bibinfo {pages} {332} (\bibinfo {year} {1992})}\BibitemShut {NoStop}%
\bibitem [{\citenamefont {{Granichin}}\ and\ \citenamefont
  {{Amelina}}(2015)}]{Granichin_TAC2015}%
  \BibitemOpen
  \bibfield  {author} {\bibinfo {author} {\bibfnamefont {O.}~\bibnamefont
  {{Granichin}}}\ and\ \bibinfo {author} {\bibfnamefont {N.}~\bibnamefont
  {{Amelina}}},\ }\href@noop {} {\bibfield  {journal} {\bibinfo  {journal}
  {IEEE Transactions on Automatic Control}\ }\textbf {\bibinfo {volume} {60}},\
  \bibinfo {pages} {1653} (\bibinfo {year} {2015})}\BibitemShut {NoStop}%
\end{thebibliography}%
